\documentclass[aps,prl,twocolumn,superscriptaddress,floatfix,tightenlines,notitlepage,nobibnotes]{revtex4}
\usepackage{graphicx}
\usepackage{bm}
\usepackage[normalem]{ulem}
\usepackage{amsfonts}
\usepackage{color}
\usepackage{verbatim}
\usepackage{amsmath}    
\usepackage{epsfig}
\usepackage[caption=false]{subfig}  
\usepackage[breaklinks,colorlinks = true,linkcolor = blue,urlcolor=blue,citecolor=blue]{hyperref}
\usepackage{amssymb}
\newcommand{\sid}[1]{\textcolor[RGB]{0,0,0}{#1}}
\newcommand{\rev}[1]{\textcolor[RGB]{0,0,0}{#1}}

\newcommand{\icts}{International Centre for  Theoretical Sciences, Tata Institute of Fundamental Research,  Bangalore, Karnataka 560089, India}
\newcommand{\iitk}{Department of Mechanical Engineering, Indian Institute of Technology Kanpur, Kanpur, Uttar Pradesh 208016, India}

\begin{document}
\title{Turbulence-Induced Fluctuating Interfaces in Heterogeneously-Active Suspensions}
\author{Siddhartha Mukherjee\textsuperscript{\S}}
\thanks{Corresponding author: smukherjee@iitk.ac.in}
\thanks{\\ \S These authors contributed equally to this work.}
\affiliation{\iitk}
\author{Kunal Kumar\textsuperscript{\S}}
\email{kunal.kumar@icts.res.in}
\affiliation{\icts}
\author{Samriddhi Sankar Ray}
\email{samriddhisankarray@gmail.com}
\affiliation{\icts}

\begin{abstract}

We investigate the effects of heterogeneous (spatially varying) activity in a hydrodynamical
	model for dense bacterial suspensions, confining 
	ourselves to experimentally realizable, simple, quenched, activity patterns.
	We show that the evolution of the \textit{bacterial velocity field}
	under such activity patterning leads to the emergence of \textit{hydrodynamic interfaces} separating spatially localized turbulence from jammed frictional surroundings. 
	We characterise the intermittent and multiscale fluctuations of
	this interface and also investigate how heterogeneity influences mixing via the residence times of Lagrangian tracers. 
	This work reveals how naturally occurring heterogeneities could decisively steer active flows into more complex configurations than those typically studied. Apart from curious parallels to droplet dynamics, front propagation and turbulent mixing layers, activity heterogeneities also present a viable route to locally controlling active flows. 
\end{abstract}

\maketitle

\section*{Significance Statement}
Self-organization into active flows in dense suspensions of living agents mediates vital biological functions like cell morphogenesis, growth and bacterial swarming. Activity is highly susceptible to variations like nutrient density gradients or light and shade patterning. While there is growing interest in heterogeneously active matter, understanding its hydrodynamical consequences remains a challenge. Making this our focal point, we show how heterogeneous activity gives rise to fluctuating hydrodynamic interfaces between coexisting turbulent and jammed frictional flows. Turbulence drives interfacial fluctuations causing entrainment, intermittent mixing and multiscaling. Our simple framework also leads to interesting parallels with front-propagation in cloud and droplet dynamics, while showing how heterogeneities can effectively localize active turbulence - a crucial precursor to engineering living materials.

\section*{Introduction}
Self-organized motion of dense active
matter~\cite{marchetti2013hydrodynamics,be2019statistical,aranson2022bacterial}
shows a range of dynamical states where relatively simple underlying
rules---the movement and interaction between active
agents~\cite{toner1998flocks,toner2005hydrodynamics} like living cells or even
externally driven inanimate
units~\cite{kumar2014flocking,nishiguchi2015mesoscopic,ginot2018aggregation}---can
end up propelling ``living'' fluids into synchrony, vortices, chaotic flows and
turbulence~\cite{sanchez2012spontaneous,wensink2012meso,dunkel2013fluid,doostmohammadi2016stabilization,doostmohammadi2017onset,chen2017weak,martinez2019selection,liu2021viscoelastic,alert2022active,james2018turbulence,james2018vortex,james2021emergence}.
Bacterial suspensions, exhibiting \textit{active turbulence}, further defy categorization by 
allowing for a range of possibilities. 
Of particular interest is the discovery~\cite{mukherjee2023intermittency} 
of a critical activity threshold $\alpha_c$  
 marking the transition from a non-universal, non-intermittent phase of chaotic flows~\cite{wensink2012meso,bratanov2015new,cp2020friction} to 
  a \textit{truly} turbulent phase~\cite{rorai2022coexistence,mukherjee2023intermittency,kiran2024onset} where the tell-tale markers associated with \textit{inertial}, high Reynolds turbulence---intermittency, non-Gaussianity of velocity moments, multiscale chaos and most importantly a universal scaling form for the energy spectrum---are manifest. This transition also marks the emergence of local vortex ordering~\cite{kashyap2025emergence}, and a surprising 
  shift from a simple diffusive system to one with anomalous diffusion~~\cite{ariel2015swarming,mukherjee2021anomalous,gautam2024harnessing} with consequences 
  for foraging and single agent dynamics~\cite{singh2022lagrangian}. 

A key simplification in all such studies of active turbulence has been the use of a \textit{constant} activity parameter $\alpha$.
Indeed, our understanding of active flows, so far, rests largely on homogeneous
living fluids: Fluids with a uniform source of nutrients and active agent
density (as in a dense bacterial suspension or microtubule-kinesin mixture),
which in continuum models for active fluids naturally translates to a uniform active
energy injection over space and time. Even such a homogeneously active fluid
can readily undergo transitions in flow states from  turbulent to coherent or
periodic vortex reversals due to
confinement~\cite{wioland2016directed,wu2017transition,chandragiri2020flow,nishiguchi2024vortex},
or become jammed by substrate friction~\cite{thampi2014active}. 

Thus it is reasonable to assume that all this dynamical variety in a homogeneously 
active living fluid might
merely be a small glimpse into a much more exotic world of active suspensions. 
After all, inevitably occurring heterogeneities --- because of
variation in light and nutrients in any environment ---  would translate to
an \textit{internal} spatio-temporal variation in the degree of activity of the living fluid, leading to modification or even inhibition of its flowing states. Consequently, such \textit{heterogeneously active} flows ought to be
more complex than their homogeneous counterparts.  Examples abound where
heterogeneities factor in, and often decisively. For instance, subtle
variations in light and shade can coax phototrophic cells to seek out
favourable illumination and non-phototrophs to evade
it~\cite{hoff2009prokaryotic,wilde2017light}, as in the case of cyanobacteria,
and exposure to intense light can locally quench collective bacterial
motion~\cite{yang2019quenching}. Similarly, nutrient and oxygen gradients at
the microscale~\cite{thar2003bacteria} drive
chemotaxis~\cite{colin2021multiple} and flagella enhanced flow
transport~\cite{petroff2015fast}. Mixed-species bacterial swarms, surprisingly,
remain heterogeneous via local segregation despite being part of a single
growing colony~\cite{natan2022mixed}. The natural course of active flows,
furthermore, must negotiate the uncertainties of obstacles and constrictions
that comprise physical environments like salt marshes~\cite{petroff2015fast}
and porous earth~\cite{engelhardt2022novel}.  Activity and confinement also
compete and a profusion of obstacles can rectify bacterial turbulence into
stable vortex
lattices~\cite{nishiguchi2018engineering,reinken2020organizing,reinken2022ising}.
 Gradients and patterns in
activity, moreover, can act as additional driving forces which may be key to
controlling active flows at will~\cite{partovifard2024controlling,shankar2024design}. They have, for instance, been shown to act as
local electric fields that can sort topological
charges~\cite{shankar2019hydrodynamics}. Optical control to shape active matter
and autonomous metamaterials by tuning activity is
burgeoning~\cite{palacci2013living,schuppler2016boundaries,ross2019controlling,zhang2021spatiotemporal,zhang2021autonomous,shankar2022topological,lemma2023spatio},
including taming bacterial motility and
density~\cite{arlt2018painting,frangipane2018dynamic,arlt2019dynamics}. Understanding the nature
of bacterial turbulence organization under heterogeneously varying activity,
therefore, is an essential step forward.

\begin{figure*}
	\includegraphics[width=1.0\linewidth]{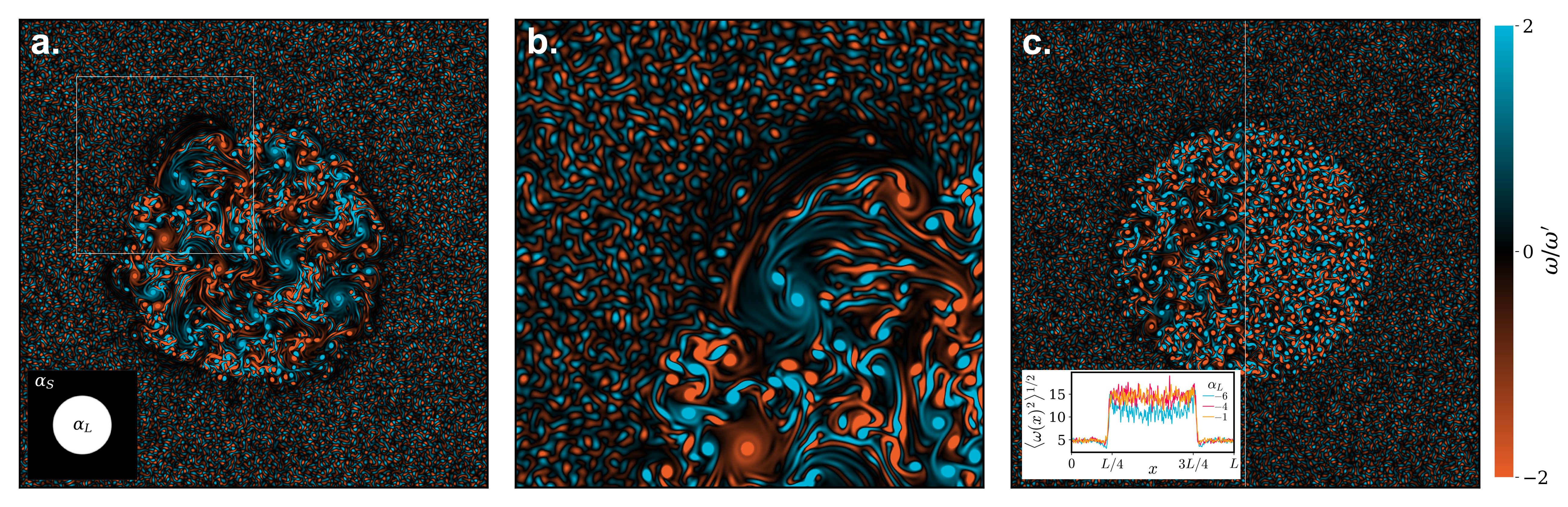}
	\caption{\textbf{Emergent interface in a heterogeneously active suspension} (a) Representative snapshots of the vorticity field $\omega$ due to a quenched activity pattern over a circular geometry, as shown in the inset (made with Processing~\cite{reas2007processing,pearson2011generative,shiffman2024nature}). There is striking coexistence of a highly active patch of turbulence corresponding roughly to the light ($\alpha_L = -6$) region that remains suspended in a frictional flow in the shadowed ($\alpha_S = 4$) fluid. This is seen even more clearly in the magnified segment shown in (b), which corresponds to the white square in (a). We find an \textit{interface} of very low vorticity that tends to separate out the active and passive flow regions. Panel (c) shows that this effect persists even upon lowering $\alpha_L$ to $-4$ (left half) and $-1$ (right half), although the interface becomes slightly sharper and does not show the large bulges observed in (a). The time evolution of the vorticity field can be seen in~\cite{omega-YT-circle}, where these effects become more evident. The inset in (c) shows the axial root-mean-squared vorticity along a diametric line passing through the active disk, averaged over time, for the typical vorticity values encountered in the two regions.}
\label{fig:IC}
\end{figure*}

While the biological consequences of heterogeneity are evidently intriguing,
the hydrodynamical aspects themselves pose challenges that remain to be
explored. With our aim to focus on the hydrodynamics, we adopt an approach based on the continuum framework of
generalized, incompressible hydrodynamics of the mean velocity field
${\bf u}({\bf x},t)$ for dense bacterial suspensions~\citep{wensink2012meso} confined to two-dimensions. We take the first steps towards systematically deviating from
familiar, homogeneously active suspensions (and hence homogeneous active
turbulence) to situations where the activity $\alpha$, and hence turbulence, varies over space. 

\rev{We work here with the widely employed
Toner--Tu--Swift--Hohenberg model, given below (see Methods for a discussion of
its parameters and experimental parallels):}

\begin{equation}
\partial_t{\bf u} + \lambda {\bf u}\cdot {\nabla\bf u} =-{\bf
	\nabla}p-\Gamma_0\nabla^2{\bf u}-\Gamma_2\nabla^4{\bf u}-(\alpha +
\beta|{\bf u}|^2){\bf u}. 
\label{GNS} 
\end{equation}

In this work, we fix $\alpha$ to have a pre-specified (quenched) spatial
pattern which remains unchanged in time: $\alpha \equiv \alpha({\bf x})$ is a
function of space and not time. This leads to uncovering coexisting turbulent
and frictional flows, separated by hitherto undetected emergent hydrodynamic
interfaces exhibiting intriguing dynamical behaviour. \rev{Here we focus on elucidating the generic hydrodynamic consequences of spatially heterogeneous activity in dense active suspensions.}

\rev{Before we proceed further with this model, it is important to add 
a caveat regarding the hydrodynamic description above. Although bacteria swim in a surrounding fluid, our approach
adopts a coarse-grained description appropriate to the high-density limit of
\textit{dense} bacterial suspensions, the regime in which these systems exhibit
turbulence and where incompressibility of the bacterial velocity field is a
reasonable approximation (particle-based models show that density fluctuations
are below $5\%$ in this regime~\cite{wensink2012meso}). A second key assumption
is that the coarse-grained bacterial orientation field is locally aligned with
the velocity, leading to a single order parameter ${\bf u}({\bf x},t)$, while
small-scale misalignments may occur in practice. Together, these assumptions
lead to an effective ``dry'' description, in which the solvent dynamics and
momentum conservation of the surrounding fluid are not explicitly retained at
the continuum level.}

\rev{This approximation is expected to be most appropriate for
dense, quasi-two-dimensional bacterial suspensions, where strong confinement or
substrate friction suppresses long-ranged solvent-mediated hydrodynamic
interactions and the flow is dominated by collective bacterial motion. By
contrast, dilute suspensions can exhibit swarming or flocking states with
strong density variations~\cite{gautam2024harnessing} and significant
solvent-mediated effects, relevant to phenomena like gyrotaxis and
bioconvection~\cite{bees2025emergent}, and concentration-wave driven
turbulence~\cite{jain2024inertia}. In the context of bacterial suspensions in
such regimes, more detailed ``wet'' models that separately track bacterial
density, velocity, and orientation fields alongside the solvent flow are more
appropriate. Alternative models distinguishing polar order and velocity fields
are also employed to study active turbulence mediated by
defects~\cite{aditi2002hydrodynamic,marchetti2013hydrodynamics,doostmohammadi2018active,ramaswamy2019active,rana2024defect,jain2025instabilities}.
A detailed comparison between dry and wet active-matter descriptions, including
their implications for active turbulence, can be found in
Ref.~\cite{alert2022active}.}

\section*{\textbf{Results}}
\begin{figure*}
	\includegraphics[width=\linewidth]{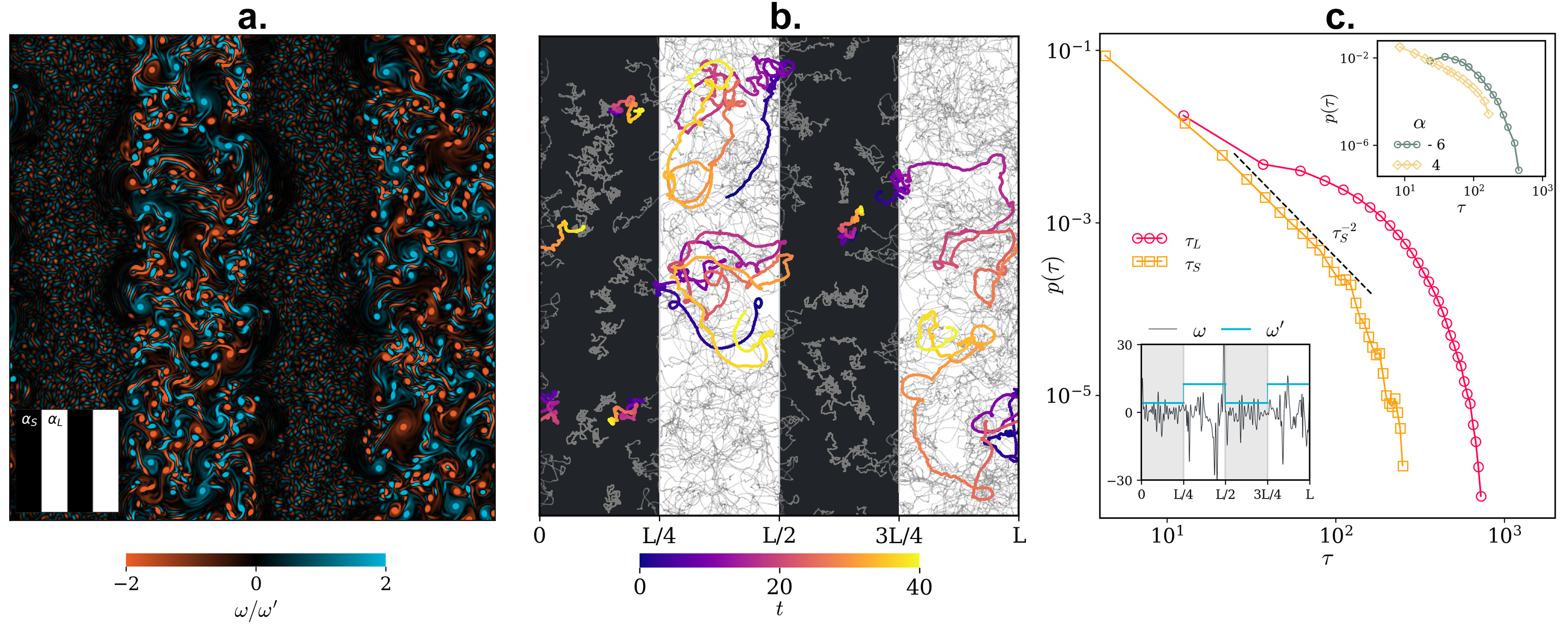}
	\caption{\textbf{Mixing across interfaces} (a) A snapshot of the vorticity field $\omega$ due to a striped activity quench (see inset), with $\alpha_L = -6$ and $\alpha_S = 4$. The vorticity fields show patterns and fluctuations across an emergent interface analogous to what was seen for the circular geometry in Fig.~\ref{fig:IC}(a), seen clearly from a movie~\cite{omega-YT-strip}. (b) Representative trajectories of several randomly selected Lagrangian particles on such a striped activity background. A subset of these trajectories are highlighted and coloured by time to contrast the short wriggly paths in the dark regions from the long persistent motion in the light regions, along with relaxation to this behaviour upon crossovers. (c) Loglog plots of the PDFs of the normalised residence times in the two strips. (Upper Inset) The analogous plots for a uniform activity with $\alpha = -6 \lesssim \alpha_c$ and $\alpha = 4 = \alpha_S$ clearly do not distinguish the artificial segmenting of the domain (see text). (Lower Inset) A plot of $\omega$ (jagged, black line) as a function of the horizontal $x$ direction and the mean value (see text) of the root-mean-square vorticity $\omega^{\prime}$ (thick blue line) in the light and shadow strips individually, showing the clear jumps as the activity strips (indicated by vertical lines) are crossed.} 
\label{fig:lagstatistics}
\end{figure*}

\paragraph*{\textbf{An emergent hydrodynamic interface}}
As a
further simplification, we limit ourselves to circular (Fig.~\ref{fig:IC}(a), inset)
or rectangularly striped (Fig.~\ref{fig:lagstatistics}(a), inset) \textit{patches} of activity $\alpha_L < 0$
surrounded by a background (frictional) activity $\alpha_S > 0$; we choose a
tan-hyperbolic interface to separate the two over a short distance of $\approx 2\%$ of the lateral extent of the physical domain. Such configurations are
experimentally viable: It has already been demonstrated to work well for
light-sensitive motor proteins powered active
fluids~\cite{zhang2021spatiotemporal}. Hence our choice of subscripts $L$
(light) and $S$ (shadow) suggests experiments on photosensitive organisms: The
more active agents, with $\alpha_L$, are confined to patches with higher light
intensity and a more frictional neighbourhood, with $\alpha_S$, shaded from the
light. 

While $\alpha_L < 0$ is the reasonable choice to locally produce turbulence,
the value of $\alpha_S>0$ needs further qualification. Of course, in the TTSH
model $\alpha>0$ and $\alpha<0$ are both permissible, and their effect on the
emerging flow has been well
studied~\cite{bratanov2015new,james2018vortex,joy2020friction,mukherjee2021anomalous,mukherjee2023intermittency}.
To create turbulence locally in a quiescent background, the analogous classical
Newtonian fluid
approach~\cite{mazzino2021unraveling,alexakis2023far,matsuzawa2023creation} is
to treat the background to be a ``still'' fluid medium with no motion of its
own. However, under the coarse grained hydrodynamics of heterogeneously active
suspensions, and what is easily conceivable in physical situations as well, one
can have coexisting flow states due to a highly motile patch (with $\alpha_L$,
which may be strongly turbulent) being immersed in a fluid that is only weakly
motile ($\alpha_S$, which we refer to as passive). There is, however, a caveat
that due to the intrinsic limitation in pushing $\alpha_L$ to values $\ll 0$
(i.e. to stay in the turbulence regime~\cite{james2021emergence} and avoid
forming condensates), a large separation in the activity of the $\alpha_L$ and
$\alpha_S$ patch cannot be achieved if $\alpha_S \lessapprox 0$ as well. We
therefore resort to treating the surrounding fluid as a frictional medium with
$\alpha_S \gg 0$, which allows $\alpha_L \ll \alpha_S$ and leads to a clear
distinction between phases. Owing to the destabilizing effect of $\lambda$ and
$\Gamma_0$, the frictional region also exhibits weak flow patterns, but as we
shall show later it may be considered quiescent in comparison to the turbulent
region. This makes the situation qualitatively closer to inertial turbulence
studies where turbulence is localized in a quiescent
fluid~\cite{alexakis2023far,matsuzawa2023creation}. Of course, we also present
a range of cases with different levels of localized turbulence immersed in the
frictional medium. More complex alternatives, we believe, will not
qualitatively change the results.

The hydrodynamics of bacterial suspensions, described by Eq.~\eqref{GNS}, allow
non-local interactions in the vorticity field (see the Methods section on details of the 
Direct Numerical Simulations (DNSs)). Consequently, even when the
activity parameter remains confined to different values in the \textit{shadow}
and \textit{light} regions, the flow field in different spatial regions are
coupled. This leads to unanticipated dynamics of the flow, as most readily seen
from the vorticity field $\omega ({\bf x}) = \nabla \times {\bf u}(\bf x)$. We begin
by taking a bird's eye view of this effect, in Fig.~\ref{fig:IC}, through
representative snapshots of the vorticity field in statistical steady states
corresponding to the circular $\alpha_L$ geometries (Fig.~\ref{fig:IC}(a),
inset). Each field is normalized by its root-mean-square vorticity $\omega^\prime = \langle \omega^2 \rangle^{1/2}$ where $\langle \cdot \rangle$ denotes spatial averaging. Figure~\ref{fig:IC}(a) shows a flow with the strongest contrast between
$\alpha_S$ ($4$, frictional) and $\alpha_L$ ($-6$, highly active).  A clear,
geometrically confined region of active turbulence, roughly corresponding to the
circular $\alpha_L$ patch, seems to emerge and persist within a jammed
frictional background. A close-up of a section of this flow (denoted by the
square perimeter, outlined in white), shown in Fig.~\ref{fig:IC}(b), is revealing: 
A hydrodynamic \textit{interface} of
near-zero vorticity separates the coexisting turbulent flow region ($\alpha_L$) from the friction flow ($\alpha_S$), 
while $\omega ({\bf x})$, deep in the \textit{lit} or \textit{shadow} regions, has
features deriving from the structure of vorticity fields we associate with the
\textit{local} value of the activity alone. We note that an imposed activity gradient seems
sufficient to behave almost like a physical, albeit deformable, boundary to the flow. Hence, this configuration
is also capable of exhibiting giant-vortex and binary vortex-pair formation (see movie~\cite{omega-giant}), as was shown for highly active turbulence under circular confinement~\cite{puggioni2022giant}. 

The emergence of an interface recurs even upon reducing the level of activity
in the light patch, as shown in Fig.~\ref{fig:IC}(c) for $\alpha_L = -4$ (left
half) and $\alpha_L=-1$ (right half). The essential qualitative change is in
the extent to which the turbulence in the light region \textit{spills} out
into the shadow region, and consequently the (dynamical) thickness of the interface.  All
these effects appear most evidently from a movie~\cite{omega-YT-circle} of the
evolution of the vorticity fields. The axial root-mean-squared vorticity $\langle \omega(x)^2 \rangle^{1/2}$ as a function of $x$ and averaged over time in Fig.~\ref{fig:IC}(c), across a diametric
line through the active disk, shows $\omega^\prime$ in the shadow regions
is $\approx 5$, while in the light region it is $\approx 14$ (with a slightly lower value of $\approx 11.5$ for the $\alpha_L = -6$ case, due to the appearance of the \textit{streaks} and \textit{voids}~\cite{singh2022lagrangian}, both of which reduce $\omega^\prime$).

The non-local interactions of the governing equations make what happens around
the interface curious and intriguing. We observe, in Figs.~\ref{fig:IC}(a) and
(c) (as well as in the movies~\cite{omega-YT-circle}), that the highly active
$\alpha_L$ patch becomes populated with a few large vortices, and myriad
smaller vortex clusters and
streaks~\cite{mukherjee2021anomalous,singh2022lagrangian,puggioni2022giant}.
The motion of the energetic large vortices, however, is not free from hindrance
as they collide with a frictional neighbourhood of vorticity outside the highly
active patch.  This leads to an arrested motion which manifests in mild
\emph{oscillations} of the interface about the prescribed $\alpha_L$ profile (Figs.~\ref{fig:IC}(a), inset),
with bulges and valleys. We observe that these vortices, and consequently the
convex bulges of the interface, tend to slowly circumscribe the circular patch
before dissipating, and there are frequent vortex pair ejections from the
\textit{light} into the \textit{shadow} region. All of this is compelling
reason to investigate the rich dynamics and fluctuations of this emergent interface more
carefully.

\paragraph*{\textbf{Undulating Height Field: Bulges and Valleys}} 
\sid{In the absence of an objective definition of the location of the interface, we develop an algorithm (see Methods section for details) to define a quantifiable contour. To this end, we exploit the fact that the turbulent flow region contains all the kinetic energy, while the passive flow region damps it to negligible values. Hence, we identify the radial location $r_\theta$ for each angle $\theta$ around the center of the circular patch, $\mathcal{C}(r_{\theta})$, where the kinetic energy $E$ of the flow crosses a threshold value. The evolution of the interface in time is denoted by $\mathcal{C}(r_{\theta},t)$. Studies on turbulent/non-turbulent interfaces in high Reynolds inertial turbulence use a similar procedure to find the interface, often using the enstrophy $\omega^2$ as the order parameter (which does not work in our case when $\alpha_S$ itself has values leading to mild active turbulence in the surrounding region, while kinetic energy works robustly).}

\begin{figure*}
	\includegraphics[width=1.0\linewidth]{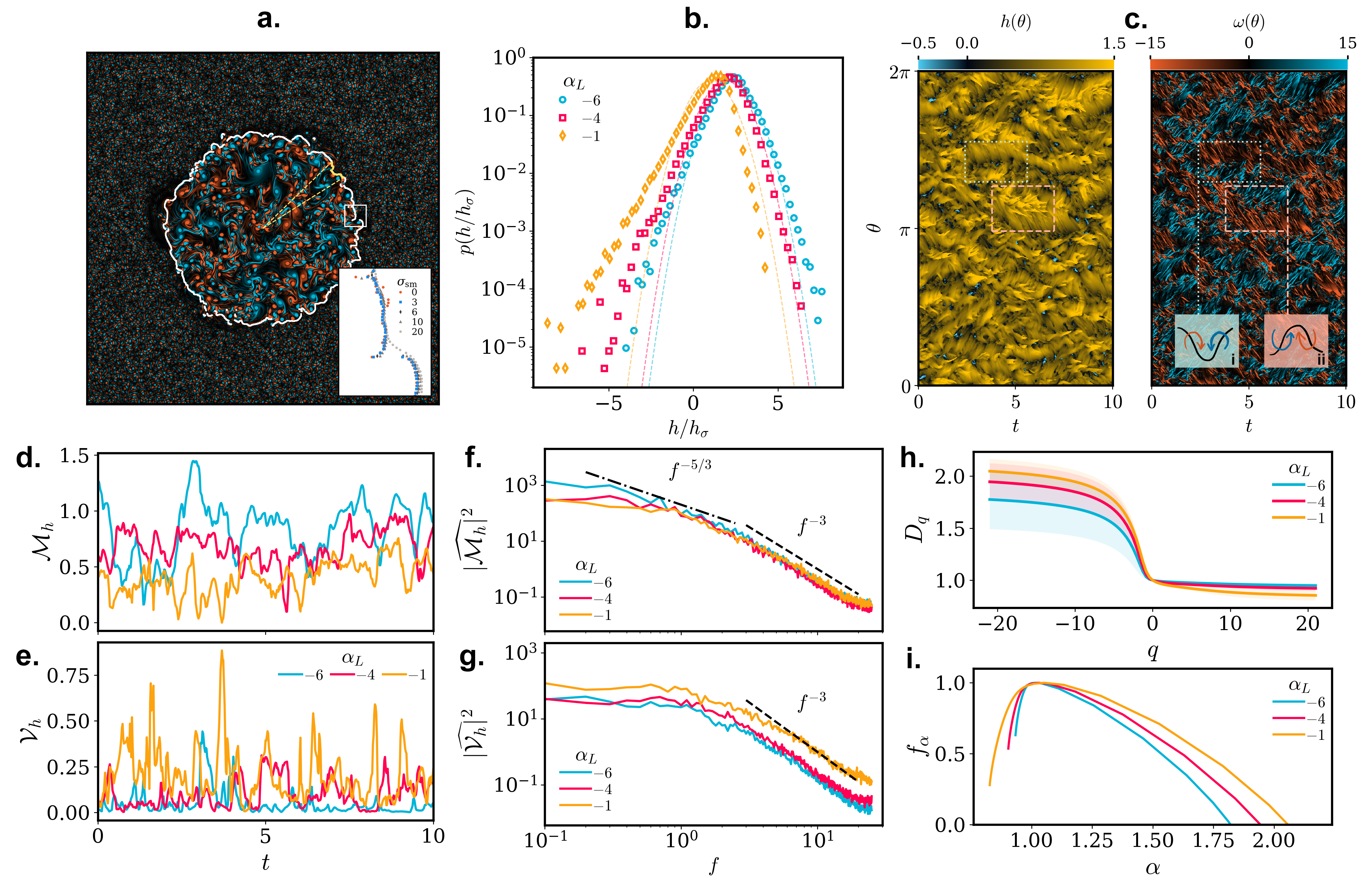}
	\caption{\textbf{An intermittent, multifractal, fluctuating interface} (a) The contour $\mathcal{C}(\theta)$ of the \textit{wiggly} interface 
	separating the turbulent flow region from the frictional surroundings, superimposed on the background vorticity field. The bottom-right inset shows a magnification of the white square, with the interface calculated using multiple smoothing ($\sigma_{\rm sm}$) values applied to the kinetic energy field. The pair of yellow lines is a guide to the eye to show how the interface is segmented for analysis. The dynamics of this interface is best seen in a movie~\cite{interface-YT-circle}. 
	(b) The probability density function (PDF) of the normalized height field $p(h/h_\sigma)$, for various $\alpha_L$; the dashed-curves are separate Gaussian fits to each PDF profile.
	(c) \textit{Space-time} plots (kymograph) of the interfacial height $h(\theta,t)$ and vorticity $\omega(\theta)$, shown for $\alpha_L = -6$ (with $\theta$ measured counter-clockwise from the equator). The dotted and dashed rectangles mark a persistent interfacial valley and bulge, respectively, which are found to occur in regions where counter-rotating vortices collide, with distinct vortex ordering along $\theta$, as sketched in the insets of the right panel. The circumferential meandering of bulges and ridges is also evident in the diagonal bands seen in both figures. A representative time-series of the sectoral (d) mean height $\mathcal{M}_h$ and (e) variance $\mathcal{V}_h$ fluctuations, for a random sector of the interface, have been shown. Loglog plots of the power spectrum of the (f) mean height $|\widehat{\mathcal{M}_h}|^2$ and (g) variance $|\widehat{\mathcal{V}_h}|^2$, averaged over all sectors, show clear power-law behaviour	indicated by the black lines. The $\mathcal{M}_h$ spectrum intriguingly shows dual scaling with exponents similar to two-dimensional inertial turbulence. Lastly, the multifractal nature of the height fluctuations is shown via the \rev{(h) time averaged generalized dimensions $D_q$ vs $q$ (where the shaded region shows the standard deviation as errorbars), and (i) Singularity spectra $f_\alpha - \alpha$, for the three values of $\alpha_L$.}}
\label{fig:interface}
\end{figure*}

In Fig.~\ref{fig:interface}(a) we superimpose the interface thus quantified, on top of the vorticity field, where for every angle $\theta$ around the origin, we now mark the radial
distance of the interface given by $\mathcal{C}(r_{\theta})$ with a point. This
set of points separates the active and passive flow regions of the
heterogeneous suspension. \sid{Since the interface calculation involves a minor smoothing applied to the base field, we demonstrate the effect of this smoothing in the inset of Fig.~\ref{fig:interface}(a) (see Methods for details). The results presented hereafter use a fine-tuned value of the smoothing parameter, $\sigma_{\rm sm} = 3L/N$, which only helps filter out the few, isolated jumps in the interface calculation while preserving the physical fluctuations.}

Importantly, although the $\mathcal{C}(r_\theta,t)$
roughly preserves the circular geometry of the quenched activity disk
(Fig.~\ref{fig:IC}(a) inset), its \textit{wiggly} nature (seen clearly in the
movie of the interface~\cite{interface-YT-circle} or the vorticity
field~\cite{omega-YT-circle}) underlines the oscillating nature of this
interface. Thence, the simplest way to understand these fluctuations is through
the statistics of the difference between the contour $\mathcal{C}(r_\theta,t)$
(Fig.~\ref{fig:interface}(a)) and the (fixed) radius of the active circular
disk $r_{\alpha_L}$ (Fig.~\ref{fig:IC}(a), inset). We do this by defining a
height field $h(\theta, t) = \mathcal{C}(r_\theta,t) - r_{\alpha_L}$, where
$r_{\alpha_L}$ is the radius of the active disk. With this definition, we note,
that $h(\theta,t)$ oscillates around a non-zero mean value since
$\mathcal{C}(r_\theta,t) > r_{\alpha_L}$, on average.

We first quantify the height field with the normalised probability density
function $p(h)$ as shown in Fig.~\ref{fig:interface}(b),
considered for all points on the interface ($\approx 4000$) and over 200 
snapshots well separated in time. 
The distributions are clearly negatively skewed: While their core
and positive tails are mostly Gaussian (indicated by the dashed lines), the
negative tails seem to decay exponentially, showing significant deviations.
Intriguingly, this is reminiscent of the distribution of pressure in inertial
turbulence~\cite{pumir1994numerical,cao1999statistics}. These negative deviations
increase as the activity is reduced in the patch, showing that the interface
surrounding mildly active turbulence suffers more intermittent ingress due to the surrounding frictional flow, 
leading to larger negative fluctuations of $h$. 
Furthermore, the mean of all the distributions is positive which
confirms the suggestion of bulging from the movies of the vorticity
fields~\cite{omega-YT-circle,interface-YT-circle}, since the oscillations of the
interface is preferentially outward. At high levels of activity, the
interfacial ingress is suppressed, and the positive tail of the distribution
also begins to mildly deviate from Gaussianity at large $h$ values.

We turn again to visualization for a clue to decipher the dynamics, in this case the 
source of the interfacial bulges and valleys, by comparing the 
interfacial height profile $h(\theta)$ alongside the interfacial vorticity $\omega(\theta) = \omega(\mathcal{C}(r_\theta,t))$.
In Fig.~\ref{fig:interface}(c) we show a space-time plot (a
kymograph)---with the vertical axis $\theta$ (measured
counterclockwise from the equator) and horizontal axis time $t$
--- of the height $h(\theta)$ and the interfacial vorticity $\omega(\theta)$ fields for the flow with $\alpha_L = -6$. 
The height fluctuations are mostly positive and
large, forming broad ridges separated by negative fluctuations that form narrow
valleys. There is clear evidence also of a diagonal banding of these structures, which reflects that the interfacial bulges and valleys often meander along the perimeter of the interface, before dissipating. The vorticity kymograph also shows diagonal banding which interestingly corresponds to regions where fast-spinning, counter-rotating vortices tend to collide and jostle. 
These regions show a strong correspondence with the bulges and valleys of the height field, and a closer look at the vortex ordering is revealing. We highlight two regions, both in the $h(\theta)$ and $\omega(\theta)$ kymographs, showing a height-field valley (dotted rectangle) and bulge (dashed rectangle). We note that negatively signed vortices rotate clockwise, and vice-versa. 
Interfacial bulges correspond to vortex pairs, encountered in the direction of increasing $\theta$, when a clockwise vortex collides with a counter-clockwise vortex, hence ejecting fluid from the light region into the shadow region, propelling the interface outward. Similarly, interfacial valleys are formed when a counter-clockwise vortex collides with a clockwise vortex, again along the interface (increasing $\theta$), which causes entrainment of the surrounding frictional flow into the turbulent light region, plunging the interface inward. We recall that the interfacial location, found based on the local value of the kinetic energy, is consistent with this reasoning as the first mechanism ejects high kinetic energy fluid (and hence $h$) outwards, while the second mechanism draws a quiescent fluid inwards (plunging $h$). These mechanisms, illustrated in the insets of the right panel of Fig.~\ref{fig:interface}(c) (where the black curve denotes a segment of the interfacial height, with $\theta$ increasing from right to left), are universally encountered in the growth of turbulent mixing-layers, jets and cumulus clouds. \sid{It is worth noting that a few instances of this correlation between the excursions of the height field and the local vortex ordering are visible in Fig.~\ref{fig:interface}(a). However, we refrain from highlighting these anecdotally, since the interface is rapidly evolving and a single snapshot can be misleading. The approach in Fig.~\ref{fig:interface}(c) is more robust, as it captures the persistence of this correlation both over space and time, albeit visually. A statistical quantification via joint-correlations will reveal as much, but such an investigation falls outside the aims of this work.}

Finally, we note that for $\alpha_L \geq -4$ (not shown here), the bulges simply persist horizontally,
showing that they do not traverse along the interfacial perimeter at mild
activity. The transition between these two qualitatively distinct behaviours again occurs
at around a critical value of $\alpha_L \leq \alpha_c$ with $\alpha_c \approx
-5$, as seen in previous
studies~\citep{mukherjee2021anomalous,mukherjee2023intermittency,kiran2024onset}.
These dynamical effects are clearly seen from the evolution of the vorticity fields and the interface~\cite{omega-YT-circle,interface-YT-circle}, and a more detailed analysis shall be done elsewhere.

\paragraph{\textbf{Two-dimensional Turbulence of Interfacial Excursions}}
We turn now to quantifying the fluctuation timescales of the interface, and find it useful to further segment the interfacial contour $\mathcal{C}(r_\theta, t)$
in sectors of angular increments ${\rm d}\theta = 2\pi/N_{\mathcal{S}}$. For illustration, one such sector and
its associated interfacial segment (for $N_{\mathcal{S}} = 32$), is shown with
dashed yellow lines in Fig.~\ref{fig:interface}(a). This segmentation allows us
to define the sectoral mean interfacial height $\mathcal{M}_h \equiv (1/N)\sum_{\theta \in \mathcal{S}} h(\theta)$ and variance $\mathcal{V}_h \equiv (1/N) \sum_{\theta \in \mathcal{S}} (h(\theta) -
\mathcal{M}_h)^2$, where $N$ is the number of points in each sector. A suitable value of $N_{\mathcal{S}}$ allows
us to quantify the average height and fluctuation of the interface over an
approximate lengthscale comparable to that of the emerging bulges, and to then
track it over time. Admittedly, the precise value of $N_{\mathcal{S}}$ is ad-hoc,
but changing it by even a factor of 2 either way does not change our findings.
For the temporal statistics that follow, we use a high resolution dataset, with
500 evenly-spaced time snapshots acquired in the statistically steady state.

 The time series of $\mathcal{M}_h$, for a representative segment, is shown in
Fig.~\ref{fig:interface}(d). Note the large fluctuations in the mean height as
a function of time for all values of $\alpha_L$. The highest activity (in
blue) also shows a lower frequency oscillation reflecting the slowly
turning bulges. The sectoral variance $\mathcal{V}_h$ in Fig.~\ref{fig:interface}(e) also
shows large fluctuations, except that the variance is larger for the
\textit{weakest} activity ($\alpha_L = -1$), reflecting the coupled effect of a pronounced
interfacial ingress (larger intermittency) and a lack of persistent bulges (smoothing out fluctuations) at low activity.
We use these high-resolution time-series to then
compute the frequency spectra averaged over all $N_{\mathcal{S}}$
sectors. Curiously, the spectrum $|\widehat{\mathcal{M}_h}|^2$ in
Fig.~\ref{fig:interface}(f) shows a clear power-law distribution with two
scaling regimes. Towards lower frequencies we find an approximately
$f^{-5/3}$ while the higher frequencies decay as $f^{-3}$. This is surprisingly 
reminiscent of the dual-cascade energy spectrum of two-dimensional
inertial turbulence~\cite{boffetta2012two}. This finding is vexing and we can only conjecture why 
this effect may be reflective, \textit{locally} on the annulus, 
of two-dimensional inertial turbulence unlike what is 
seen deep inside the light or shadow regions (with their own distinct flows).
The highly active disk, specially for the $\alpha_L \lesssim \alpha_c$ creates an inertial flow which is
flung outwards into frictional surroundings as well as drawing quiescent flow suddenly inwards. 
This situation is similar to two dimensional inertial turbulence 
where large scale organization is met with Ekman friction. This interaction of inertia and 
an effective Ekman friction is limited naturally to the neighbourhood
around the interface leading to a ring of two-dimensional flows akin to inertial 
turbulence. Given the likely linear relationship between the height field and 
the local velocity field, it is possible that the power spectrum of the height field reflects 
the scaling of the more conventional energy spectrum in two-dimensional turbulence. 

The $\alpha_L = -6$ spectrum also
confirms low-frequency oscillations of the interfacial height, which is a
consequence of the persistent turbulent bulges arising due to the highly
active patch. The spectra of the sectoral height variance $|\widehat{\mathcal{V}_h}|^2$
Fig.~\ref{fig:interface}(g) also shows an approximate $f^{-3}$ decay at high
frequencies, more clearly for the highly active patches than for the mild
activity patch. The key finding from the spectra is that the height
fluctuations are multiscale, borne out of the multiscale nature of the
underlying flow~\cite{mukherjee2023intermittency}.

These features of the emergent interface between the active and passive flow
regions are intriguing. While fat-tailed fluctuations of $h$ are
suggestive of intermittency, the power-law decay of
$\mathcal{M}_h$ reflects a multiscale temporal structure and possible
self-similarity (which naturally would translate to multiscale spatial
fluctuations as well). Intermittency has been found robustly in homogeneously
active flows, both in spatial and
Lagrangian~\cite{mukherjee2023intermittency,kiran2024onset} measurements. It is
interesting that beyond velocity difference statistics, as reported in these
studies, even the interfacial height that separates different flow regions shows intermittent fluctuations. 
A final quantification of
the nature of these fluctuations, and their degree of self-similarity, is done
via a multifractal analysis~\cite{Frisch-Parisi} of the $h(\theta,t)$ profile.

\paragraph{\textbf{Multifractal Fluctuations}}
\rev{Here we use multifractal analysis as a diagnostic tool to establish the presence of intermittent, multiscale fluctuations in the emergent interface, rather than as an exhaustive characterization of all scaling properties.}
We follow the standard procedure for calculating the multifractal spectrum~\cite{Frisch-Book,MS-Nucl,MS-PRL,MS-JFM}, which for a $d$-dimensional field of size $L^d$, starts by summing the total measure in $d-$dimensional ``boxes'' of size $l$, denoted by say $\mathcal{H}_l$. Upon taking their $q^{\mathrm{th}}$ moment and summing over all $N_l (\approx (L/l)^d)$ boxes, one obtains the \textit{partition function} $Z_q(l) \equiv \displaystyle \sum_{N_l} \mathcal{H}_l^q \sim l^{(q-1)D_q}$. Here $D_q$ are the generalized dimensions~\citep{grassberger1983characterization} (also called the Renyi dimensions), and are found from the scaling between $\ln Z_q^{1/(q-1)}$ and $\ln l$. To get to the spectrum $f_{\alpha}$ from the calculation of $D_q$, one uses the relation that $\left\lbrace q,(q-1)D_q \right\rbrace$ and $\left\lbrace \alpha ,f_{\alpha }\right\rbrace$ forms a Legendre transformation pair~\cite{Frisch-Book,MS-Nucl} (see also SI of~\cite{mukherjee2024turbulent}), and hence the singularity spectrum is given by $\alpha = \frac{d}{dq}[(q-1)(D_q -d + 1)]$ and $f_\alpha = \alpha q - (q-1)(D_q - d + 1) + d-1$. More details can be found in the Methods section.

The generalized dimensions $D_q$ vs $q$ in Fig.~\ref{fig:interface}(h) show a
broad distribution more pronounced towards the negative $q$ values, while the
positive $q$ part of the distribution remains relatively flat.
\rev{Upon examining $D_q$ for higher-order moments, we find that
the branches of the $D_q$ curve remain essentially flat. While this might
suggest convergence, two important physical nuances clarify this behavior. The
probability distribution of the coarse-grained measure $\mathcal{H}_l$ becomes
increasingly log-normal as the coarse-graining scale $l$ increases, reminiscent
of other intermittent fields in turbulence, such as the coarse-grained energy
dissipation in three-dimensional turbulence~\cite{kolmogorov1962refinement,arneodo1998towards}. As in those cases, coarse-graining
limits the resolution of the far tails of the distribution, even though its
central structure remains well resolved. In addition, and in contrast to
three-dimensional dissipation, the interfacial height field possesses a natural
cutoff: while the interface is intermittent, it does not produce arbitrarily
extreme fluctuations (see Fig.~\ref{fig:interface}b). 
Multiscale fluctuations, and therefore multifractality, arise predominantly at moderate height magnitudes, whereas the largest deviations are possibly limited by the smoother, coherent structures of the large-scale vortices. As a result, higher-order moments become increasingly
dominated by a small number of extreme events, causing the generalized
dimensions $D_q$ to appear saturated as $q$ increases. This behavior reflects
the physical nature of the fluctuating interface rather than a lack of
statistical convergence, and a finite spread in $D_q$ persists over the range
$-10 < q < 10$, leading to a broad multifractal $f_\alpha-\alpha$ spectrum.
Additional convergence checks, based on the probability distributions of $\log
\mathcal{H}_l$ at different coarse-graining scales (not shown), confirm that the
multifractal characterization remains statistically robust over the reported
range of moments.}

The spectrum of singularity strengths $f_\alpha-\alpha$ in Fig.~\ref{fig:interface}(i) show a 
positively skewed distribution admitting a range of $\alpha$ values and hence suggesting 
that the fluctuations follow a range of scaling exponents. Interestingly, with
increasing activity in the patch the degree of multifractality reduces. The
outward bulges that dominate in the highly active $\alpha_L = -6$ turbulence
tend to smoothen the interfacial profile causing an overall reduction in the 
range of singularity exponents of the interfacial height. 
The $\alpha_L = -1$ case, thus, shows
the most multifractal fluctuations. \sid{Finally, we checked that the spread of the multifractality exponents is not influenced by the smoothening procedure employed in the interfacial calculation (see Methods).}

We arrive at a curious parallel
between this emergent hydrodynamic interface---without any \textit{real} physical
barrier separating the coexisting flow regions---with the fluctuating interfaces
of droplets in inertial turbulence~\cite{pal2016binary} and the self-induced
fluctuations of an active droplet in a passive fluid
medium~\cite{padhan2023activity,padhan2025cahn}: Both these cases of immiscible fluids separated by
an elastic interface manifest multifractal fluctuations. 

\paragraph{\textbf{Transport and Mixing}} The emergence of intermittency in the structure of the flow field was shown
earlier to have important consequences for perhaps the more direct problem of
transport and mixing in such dense
suspensions~~\cite{ariel2015swarming,mukherjee2021anomalous,singh2022lagrangian,chamkor-chaudhuri-natcomm,kiran2024onset}.
To this end, we introduce tracers particles into the flow and monitor their
interweaving transport across differently active regions. We find studying such
Lagrangian aspects particularly useful on periodic patterns. The simplest
example of this is a striped light ($\alpha_L = -6$) and shadow ($\alpha_S = 4$)
array which creates a periodic pattern with equal regions of active and
passive flow. This greatly facilitates the comparison of Lagrangian measures,
preferentially sampled in different regions, without having to correct
additionally for the geometrical asymmetry between the light and shadow
regions. In Fig.~\ref{fig:lagstatistics}(a) we show a snapshot of such a
vorticity field (in steady state) for the striped configuration, along with an
inset showing the quenched activity pattern. Similarly to the circular
geometry, we find clearly emerging bands of highly active flow, with an
undulating interface separating frictional flow regions (see movie~\cite{omega-YT-strip}). This also confirms that the
essential features of heterogeneously active flows and emergent interfaces are robust and consistent
across different geometrical configurations.

Naturally, such a pattern allows us to effectively 
track the influence of active and passive regions on tracer diffusion and their
residence times in the coexisting flow phases. 
Figure~\ref{fig:lagstatistics}(b) shows representative trajectories of several
randomly chosen Lagrangian particles against a backdrop of the quenched
activity pattern. We further highlight a few of these trajectories to bring out
their time evolution lucidly: Small, wriggly paths in the dark regions and
long, persistent motion in the light regions.  Crossovers between regions lead
to a relaxation to these characteristic features.  This gives us clues as to
possible preferential sampling of light and dark regions --- in terms of
residence times --- of individual trajectories.
A quantitative measure of this preferential sampling is possible through a
calculation of the residence time statistics conditioned on the spatial
location of trajectories (see the Methods section of how such trajectories are calculated). 
It is worth recalling that residence time
approaches in the study of flow structures in high Reynolds number turbulence are  
not new. Previous studies~\cite{persistencePRL2011,3Dpersistence} have
used a similar approach for the lifetimes of vortical and straining
regions of a turbulent flow, as well as their role in the trapping of Lagrangian
tracers. More recently, this was studied for active turbulence~\cite{active-persistence} 
highlighting the difference between high and low Reynolds number flows.

Our problem, though, is slightly different. Given the quenched nature of the
activity pattern that we choose, the residence time in either the turbulent or frictional
flow regions reflect the slightly different --- though coupled through
non-local terms --- dynamics of the velocity field in these two regions. There
is one further subtlety in the question of residence times in heterogeneous
regions. The typical, intrinsic flow timescales for the turbulent (light) and frictional (shadow) regions
are different. A simple way to quantify this is through the inverse of the
root-mean-square vorticity in these regions. In
Fig.~\ref{fig:lagstatistics}(c), bottom inset, we show a plot of $\omega$ along
the horizontal $x$ direction, at an arbitrary $y$ location and time.
Understandably, the profile shows fluctuations, that are discernibly larger in
the light regions. This is seen clearly in a plot of the axial root-mean-square
vorticity $\omega^{\prime}(x) \equiv \langle \omega (x,y)^2 \rangle_y^{1/2}$,
as a function of $x$, with $\langle \cdot \rangle_y$ denoting averaging along
the $y$ direction. This yields mean values $\omega^{\prime}_L$ and
$\omega^{\prime}_S$, for the light and shadow regions, which have been further
averaged over time and across the width of the strips, shown in the same
figure. Thence, the strips are associated with two different intrinsic time
scales: $\tau^{\rm int}_L \sim 1/\omega^{\prime}_L$ and $\tau^{\rm int}_S \sim
1/\omega^{\prime}_S$.

Let us now return to the trajectory of an $i$-th particle. As is clear from
Fig.~\ref{fig:lagstatistics}(b), an individual Lagrangian particle travels in
and out of strips with different activities. We therefore segment the
trajectory into $\tau^\prime_L$ and $\tau^\prime_S$ intervals of time:
Repeating this process for all the trajectories allows  us to construct the
residence time probability density functions. We then factor in the intrinsic
time-scales of strips with different $\alpha$, by considering the normalized
residence times $\tau_L = \tau^{\prime}_L/\tau^{\rm int}_L$ and $\tau_S =
\tau^{\prime}_S/\tau^{\rm int}_S$.

In Fig.~\ref{fig:lagstatistics}(c) we show a loglog plot of the PDFs of the
residence times $\tau_L$ and $\tau_S$ on the two differently active strips of
our suspension for $\alpha_L = -6$ and $\alpha_S = 4$ (where we have excluded trajectories 
that never crossover from one region to the other). 
We find that for residence times conditioned on the trajectories being trapped
in the shadow region, the distribution shows a distinct power-law regime with an
$\alpha_S$-independent (for  $\alpha_S > \alpha_c$) scaling exponent $\gamma \approx -2$.
In contrast the light region ($\alpha_L < \alpha_c$) develops a distribution with a broadened peak at an intermediate $\tau_L$, with a clear exponential tail. 

In the absence of a theory for this power-law (and hence $\gamma$) or the
transition from a self-similar distribution to one with exponential tails, we
check for self-consistency in our analysis. \sid{Could the difference between these distributions be just a trivial consequence of the geometry of the pattern and not of the contrasting activities underlying such stripes? This is easily checked by 
performing two simulations with a uniform $\alpha$ but where, during analysis, the flow is 
artificially segmented in strips with the same geometry as in the inset of
Fig.~\ref{fig:lagstatistics}(a). In the top inset of
Fig.~\ref{fig:lagstatistics}(c) we show the analogous plots for a uniform
activity with $\alpha = -6 \lesssim \alpha_c$ and one with $\alpha = 4 \gg
\alpha_c$. Here too the $\alpha = 4$ ($\alpha_S$ of the heterogeneous case) exhibits a power-law and $\alpha=-6$ (corresponding to $\alpha_L$) has an exponential distribution, which affirms consistency with the main panel of Fig.~\ref{fig:lagstatistics}(c). However, for the homogeneous activity cases shown in the inset, the separation between the distributions is minimal, while that in the main panel of Fig.~\ref{fig:lagstatistics}(c) is rather pronounced, which reflects that the difference is not simply an artefact of geometry, but an effect of the non-trivial coupling between the light and shade patches borne out in the tracer residence times.} 

\sid{Naturally, the resultant mean square
displacements are also sensitive to where the tracer particles are: However, this
dependence is simply connected to what we already know for homogeneously active
suspensions and how the value of $\alpha$ --- higher or lower than $\alpha_c$ --- 
determine normal or anomalous diffusion~\cite{mukherjee2021anomalous}.
Nevertheless, such Lagrangian analyses need to be explored in more detail 
in future studies to make direct connections with fundamental biological  
strategies where individual agents could leverage the non-trivial dynamics of emergent flow interfaces
in heterogeneously active media.}

\section*{\textbf{Discussions}}

To summarise, in this work we show that activity heterogeneities, even in
simplistic settings of quenched spatial patterning, can lead to compelling
dynamical complexity involving coexisting turbulent and quiescent flow states,
and emergent fluctuating interfaces. These findings are a very timely
complement to spatio-temporal activity patterning
studies~\cite{zhang2021spatiotemporal,lemma2023spatio}, presenting the missing
hydrodynamical perspective. Furthermore, we independently parallel 
recent observations in a contemporary, like-to-like
experimental work on confined bacterial suspensions, where self-organization
leads to \textit{interfaces} separating motile shells from not-motile
cores~\cite{Patteson2018,hokmabad2025spatial,martinez2025interfacial}. 
We believe this highlights the need to study heterogeneous active
hydrodynamics in tandem with experiments. \rev{As prefaced early on, our key findings on fluctuating interfaces
are not limited to the model employed in this study. Recently, a similar setup with a nematic model revealed topological defect, and hence turbulence, proliferation with activity confinement~\cite{partovifard2026controlled}, a situation where our findings and approach to emergent interfaces will also be relevant.}
Further in this vein, our work also shows how activity
gradients can act as effectively as boundaries confining turbulent flow, even
sustaining the formation of giant vortices and binary-pairs typically found in
\textit{geometrically confined} highly active
flows~\cite{puggioni2022giant,puggioni2023flocking}. \rev{The
results presented here are expected to apply most directly to
physical systems with high bacterial cell density, operating in the
incompressible limit. Extending the study of spatially heterogeneous activity
to particle-based models, where density variations play a central role, may
reveal interesting connections to phenomena such as motility-induced phase
separation. Spatial modulation of activity is also expected to strongly
influence synchronization, collective beating, and chaotic flows in systems
driven by cilia or bacterial carpets, where long-range hydrodynamic
interactions are
essential~\cite{uchida2010synchronization,golestanian2011hydrodynamic,chakrabarti2023collective,chakrabarti2024cytoplasmic}.
In such scenarios---particularly in mixed active--passive systems exhibiting
collective turbulence---more detailed, momentum-conserving descriptions that
explicitly account for solvent-mediated interactions may be
required~\cite{peng2021imaging,martinez2021scaling}. In three-dimensional settings where flocking and turbulence can coexist~\cite{perlekar2026flocking}, activity heterogeneities are likely to decisively alter flow transitions.}

Similarly to what is found in inertial turbulence~\cite{alexakis2023far,matsuzawa2023creation}, a patch of active
turbulence also does not ``spread'' far in a quiescent background. Both these findings reinforce a simple route to engineering
isolated patches of active turbulence via heterogeneous activity.
Fluctuating interfaces, moreover, pose new problems in the study of living fluids echoing challenges encountered in \sid{high Reynolds number inertial turbulence in three-dimensions}, like entrainment in mixing layers and subsiding cloud shells, turbulent/non-turbulent interfaces ~\cite{heus2008subsiding,westerweel2009momentum,chauhan2014turbulent,watanabe2015turbulent,borrell2016properties,elsinga2019turbulent,nair2021lagrangian,Sahoo2025}, as well as problems involving turbulent front-propagation~\cite{pocheau1994scale,xin2000front,koudella2004reaction,corwin2012kardar,bentkamp2022statistical,roy2023small}.
Taking a first step in this direction, we complement our Eulerian approach with a
Lagrangian perspective of mixing and transport under activity heterogeneity and
show how preferential sampling of different flow regions emerges quite
naturally. \sid{While our results are from a two-dimensional system, we believe localized activity in three-dimensional living fluids is bound to develop similarly isolated turbulence and interfaces.}
\sid{Given the relative simplicity of our approach, we hope this study
will lead to experiments on the control and tuning of living fluids,
geared towards engineering active flows to will. This also brings us to interesting crossroads where
biologically relevant strategies like enhanced colony growth and elevated
resistance to antibiotics are possibly conferred by the emergent hydrodynamics
of heterogeneous suspensions~\cite{lai2009swarming,butler2010cell}, studying which demands recourse to more generalized forms of activity.}

\section*{\textbf{Methods}}
\sid{Within the continuum framework of generalized, incompressible ($\nabla\cdot{\bf u}=0$) hydrodynamics for a dry, polar active fluid modelling dense bacterial suspensions, the coarse-grained velocity field ${\bf u}({\bf x},t)$ evolves according to the Toner-Tu Swift-Hohenberg (TTSH) equations~\citep{wensink2012meso} given as:}
\begin{equation}
\partial_t{\bf u} + \lambda {\bf u}\cdot {\nabla\bf u} =-{\bf
	\nabla}p-\Gamma_0\nabla^2{\bf u}-\Gamma_2\nabla^4{\bf u}-(\alpha +
\beta|{\bf u}|^2){\bf u}. 
\end{equation}

The parameter $\lambda>1$ corresponds to pusher-type swimmers and the two $\Gamma$-terms along with the convective derivative result in the formation self-sustained chaotic flow patterns. The precise nature of the suspensions are really determined by the Toner-Tu drive with $\alpha < 0$ leading to local polar ordering (while	$\alpha>0$ acts as Ekman friction), while $\beta>0$ for stability. \sid{The value of the various parameters, given below in the section on Direct Numerical Simulations, are all kept the same as those originally tuned to match experiments~\cite{wensink2012meso}. In this, like in past studies, the only parameter we allow ourselves to vary is the activity $\alpha$. Bacterial suspensions have typical coarse-grained velocities in the range $25\ \mu m/s - 100\ \mu m/s$, depending upon the physical conditions of the
experiment~\cite{wensink2012meso,sokolov2012physical,ariel2015swarming}. Activity values in the range $-6 \leq \alpha \leq -1$ (with $\alpha = -6$ being higher activity than $\alpha = -1$), lead to simulation velocities that fall within the experimental range (although we note that these estimates are approximate and a direct mapping is not possible). While the original study~\cite{wensink2012meso} proposes an active injection of $\alpha = -1$, subsequent numerical studies~\cite{bratanov2015new,james2018turbulence,james2018vortex,cp2020friction,james2021emergence} have explored activity both in higher injection and frictional levels, leading to a broader understanding of the TTSH model and its various pattern-forming phases. Pushing the activity to higher levels beyond a critical $\alpha_c \approx -5$ has been found crucial in revealing a transition to universality and intermittent turbulence in the TTSH model~\cite{mukherjee2023intermittency}, with scaling exponents comparable to certain experimental studies~\cite{li2022topological} and to similar asymptotic states in active nematic turbulence~\cite{linkmann2019phase,saghatchi2022nematic,rorai2022coexistence}. Curiously, the TTSH model with $\alpha \leq \alpha_c$ also manifests anomalous superdiffusion via L\'evy walks, as experimentally found in bacterial colonies and suspensions~\cite{ariel2015swarming,gautam2024harnessing}, along with a host of Lagrangian anomalies and dynamical heterogeneity~\cite{singh2022lagrangian}.}
		
In all these past studies, the activity $\alpha$ has been considered a constant in
	space and time. But, as discussed in the Introduction section, in most conceivable,
	experimental situations this is unlikely to happen: The activity ought
	to be a function of space ${\bf x}$ and time $t$. The consequences of a
	variable $\alpha$ are an open question, and one that we answer now. 

\paragraph{\textbf{Direct Numerical Simulations}}
We perform direct numerical simulations (DNSs) of the TTSH equation with a 1/2
de-aliased pseudo-spectral algorithm to account for the cubic non-linearity on
square periodic boxes of length $L=20$ to $80$ discretized over $N^2=1024^2$ to
$4096^2$ collocation points. All parameters are kept the same as in previous
studies, i.e. $\Gamma_0 = 0.045, \Gamma_2 = \Gamma_0^3, \beta = 0.5, \lambda =
3.5$~\cite{james2018turbulence,james2018vortex,mukherjee2021anomalous,singh2022lagrangian,mukherjee2023intermittency}, \sid{as had been originally mapped from experiments}~\cite{wensink2012meso}.
We use a second-order Runge-Kutta scheme with a time-stepping ($\Delta t$) of
0.0002, and the linear terms ($\Gamma_0,\Gamma_2$) are treated with an
integrating factor while the $\lambda,\alpha$ and $\beta$ terms are calculated
in real-space. For ease of understanding the rich 
dynamics in such heterogeneous suspensions, we choose $\alpha_S = 4$ and 
$-6 < \alpha_L < -1$, and present results from circular geometries with $L=40$ and radius of the active region $r_{\alpha_L} = L/4$, and from striped geometries with $L=30$ and width of the active strips $w_{\alpha_L} = L/4$. We have checked that our results and conclusions remain robust for 
$-1 < \alpha_S < 5$ as well as for geometries beyond the circular or striped patterns 
shown in this study.

\begin{figure}
	\includegraphics[width=1.0\columnwidth]{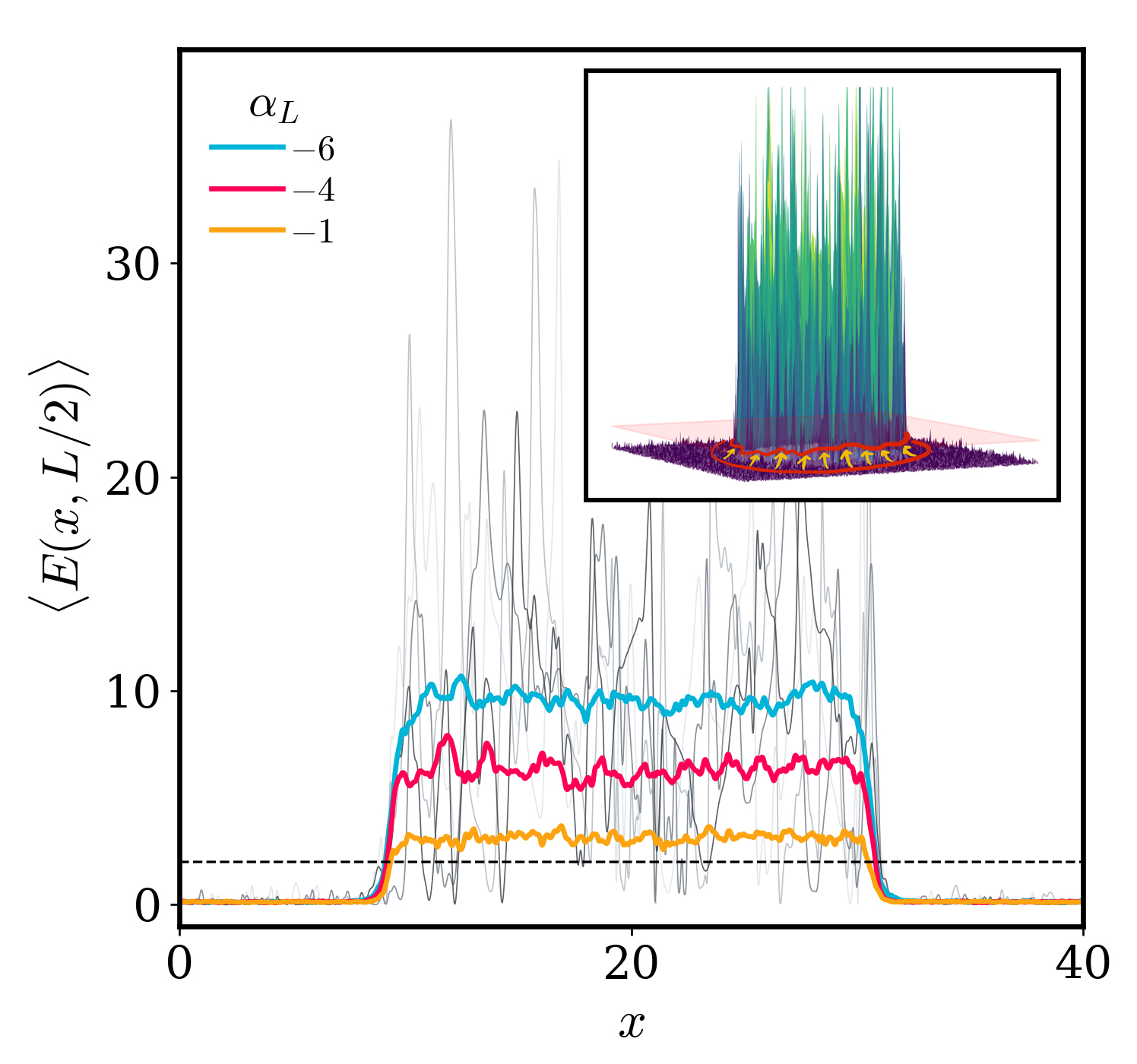}
	\caption{\textbf{How to determine an interface?} Time averaged kinetic energy $\langle E (x,L/2)\rangle$ along a diametric line through the highly active \textit{light} patch, at $y=L/2$ showing a transition in the magnitude of $E$ as one moves from the shadow to the light region. The grey lines are a few samples of the instantaneous kinetic energy along this line, for $\alpha_L=-6$, showing that the transition point fluctuates along $x$ over time. The horizontal dashed line shows the threshold we use to determine the interfacial location, $E = 2$, a value suitable for all cases. 
The inset shows a schematic of how we determine the interface---a surface plot of instantaneous $E$ is intersected by the iso-plane of $E = 2$, the interface is the outer-hull of the intersections, determined by an inward marching procedure (denoted by the yellow arrows).}
\label{fig:MeanKEDia}
\end{figure}

\paragraph*{\textbf{Defining the contour $\mathcal{C}({\bf x},t)$ of the interface}} 
It is essential, therefore, to first define in a measurable way the
location of the interface. Paradoxically, while the interface is obvious to the eye
(Figs.~\ref{fig:IC}(a) and (c)), a quantifiable measure of the interfacial
contour $\mathcal{C}({\bf x},t)$ is subtle in the absence of a natural
order-parameter separating the \textit{flow} in the light and dark patches. We
develop a simple algorithm to determine $\mathcal{C}({\bf x},t)$, using the idea that
some features of the flow \textit{transition} as one moves from the shadow to
the light region, radially inward. While the $\omega ({\bf x})$ field clearly shows a
\textit{low} vorticity annulus acting as the interface, regions of low vorticity
are also found deep in the turbulent and frictional flow regions, rendering it an
ineffective measure of the transition. The kinetic energy $E({\bf x}) = |{\bf u}(\bf x)|^2$, instead, shows a clear
separation between turbulence and its surroundings, attaining large values
preferentially in the highly turbulent region only. In
Fig.~\ref{fig:MeanKEDia} we show the time averaged kinetic energy $\langle E
(x,L/2)\rangle$, for different $\alpha_L$ (but the same $\alpha_S = 4$), taken along a diametric line through the highly
active patch and hence varying only along $x$. There is a
clear jump in the profile as one approaches the highly active region from the
outside. The grey, fluctuating lines
show a few examples of the instantaneous kinetic energy for the $\alpha_L = -6$
suspension, underlining that this transition point fluctuates in time (the inset shows the instantaneous kinetic energy as a surface, further highlighting this fact).

This gives us a tractable route. For every angle $\theta$ around the center of
the domain, also corresponding to the center of the \textit{light} region, we
define a set of initial points on a circle of radius $r > r_{\alpha_L}$ (we fix
$r=1.5 r_{\alpha_L}$). From each point, we march inwards toward the center of
the domain in step sizes corresponding to the size of the simulation grid cells
$L/N$ (smaller step sizes were also tested, but found unnecessary). We retrieve
the kinetic energy value $E(r \cos \theta, r \sin \theta)$ by considering the
position $(r \cos \theta, r \sin \theta)$ as integer multiples of the grid
length $L/N$. (A bilinear interpolation to obtain the kinetic energy at off-grid
locations was also checked, but the interfacial profiles yielded were
essentially identical, so we adopt the simpler approach.) We define the
interfacial location as $\mathcal{C}(r_\theta,t)$, for a given angle $\theta$, as
the radial distance $r_\theta$ where the kinetic energy crosses a threshold
value $E({\bf x},t) = 2$ for the first time while marching inward 
(threshold values in the range $1 < E({\bf x},t) < 3$ do not alter the results qualitatively). The reason
for picking this threshold is because it works for all three
$\alpha_L$ cases considered in this study; see the horizontal dashed line in
Fig.~\ref{fig:MeanKEDia} (and horizontal plane in the inset, while the arrows illustrate the marching algorithm to find the interface). 
Naturally, the algorithm is sensitive to fluctuations of
the underlying field, so we further test it on the filtered kinetic energy fields
$\widetilde{E}({\bf x})$, employing a Gaussian filter of standard deviation
$L/N \leq \sigma_{\rm sm} \leq 20 L/N$ (here $\sigma_{\rm sm}$ serves as a
smoothing parameter). Obviously, a larger filter begins to smoothen even
physically relevant fluctuations, and we find that $\sigma_{\rm sm} \approx 3
L/N$ gives a good balance between suppressing sudden jumps while maintaining
natural variations of the contour.

The inset in Fig.~\ref{fig:interface}(a) shows a
magnified view of the interface alone, computed on filtered kinetic energy
fields $\widetilde{E}$ with different values of the smoothing parameter
$\sigma_{\rm sm}$. We find that $\sigma_{\rm sm} \approx 3$ suppresses sudden
fluctuations present in the unsmoothed ($\sigma_{\rm sm}=0$) profile, while
still retaining the physically relevant fluctuations that get suppressed for
$\sigma_{\rm sm} > 6$. 

\paragraph*{\textbf{Obtaining the multifractal statistics}} For this, we
consider the function $\mathcal{H} (\theta) = |h(\theta)| + \gamma$ (where
$\gamma = 0.001$ is a small number added to offset the entire height profile
to be positive definite). Since $\theta$
essentially identifies $\mathcal{H}$ along the perimeter of the interface,
corresponding to a specific point on the base $\alpha_L$ profile ${\bf
x}=(r_{\alpha_L}\cos\theta,r_{\alpha_L}\sin\theta)$, we use it as a proxy for 
spatial location itself, which can be easily related as $x =
r_{\alpha_L}\theta$ or $x \propto \theta$, for ease of interpretation in the
analysis. We construct the partition function in the usual
way~\cite{frisch1995turbulence,meneveau1987simple,meneveau1991multifractal,mukherjeelocalPRL2024} as
$Z_q(l) \equiv \sum_{i=0}^{N_l} \mathcal{H}_{l,i}^q \sim l^{(q-1)D_q}$, where
$l$ is the coarse-graining length $0 < l < \mathcal{L}$, $\mathcal{L}$ is the
total length along the perimeter, $\mathcal{H}_{l,i}$ is the coarse grained
height at scale $l$ and the $i-$th partition, given as $\mathcal{H}_{l,i} =
\sum_{x=li}^{l(i+1)} \mathcal{H}(x)$ and $N_l = \mathcal{L}/l$ is the number of
partitions at lengthscale $l$. Here, $D_q$ are the generalized
dimensions~\cite{grassberger1983characterization}, where the scaling of $\ln
Z_q^{\frac{1}{q-1}}$ vs $\ln l$ gives the distribution of $D_q$ vs $q$. We
calculate the time averaged generalized dimensions, by ensemble averaging the
$D_q$ vs $q$ curve over 500 snapshots of the interfacial height, and with that
arrive at the singularity spectrum $f_\alpha - \alpha$ via a Legendre
transform, for 1-Dimensional data, as $\alpha = \frac{\rm d}{{\rm d}q}
(q-1)D_q$ and $f_\alpha = \alpha q - (q-1)D_q$. At an operational level, we truncate $\mathcal{L}$ to $4096$ points along the $\theta$ direction (hence skipping a
small part of the interface in the calculation), which has no effect on the
outcome, but allows the data to be completely tiled with $l \in \lbrace {2^0,
2^1, 2^2 ...\ 2^{12}} \rbrace$ points.

\sid{We further checked that the spread of the multifractality exponents is not influenced by the smoothening procedure employed in the interfacial calculation, for even upto twice the level of smoothening that we use (i.e. upto $\sigma_{\rm sm} = 6L/N$). Naturally, at significantly higher levels of smoothening ($\sigma_{\rm sm} \geq 10L/N$), the interface begins to lose its fluctuations, which causes the range of exponents to shrink. Such a test is imperative in the absence of a rigorous definition of the interface to establish robustness of the results from the multifractal analysis.}

\paragraph{\textbf{Obtaining particle trajectories}} We seed the flow, in a
statistically steady state, with $10^5$ tracers distributed randomly over the
whole domain. The instantaneous velocity of tracers is given by the equation
$d{\bf x}/dt={\bf u}({\bf x}(t))$, where ${\bf x}(t)$ is the tracer location.
The tracers are evolved using a fourth-order Runge-Kutta scheme, with the fluid
velocities at tracer locations estimated using bilinear interpolation.
 
\section*{\textbf{Data availability}}
The raw simulation data will be available upon request in a hard drive because of its large size (TB). The processed data of the plots are available as Source Data files.

\section*{\textbf{Code availability}}
The simulations are performed with codes developed in-house; they are available from the authors upon reasonable request.

\bibliography{references}

@article{shankar2024design,
  title={Design rules for controlling active topological defects},
  author={Shankar, Suraj and Scharrer, Luca VD and Bowick, Mark J and Marchetti, M Cristina},
  journal={Proceedings of the National Academy of Sciences},
  volume={121},
  number={21},
  pages={e2400933121},
  year={2024},
  publisher={National Academy of Sciences}
}

@article{partovifard2024controlling,
  title={Controlling active turbulence by activity patterns},
  author={Partovifard, Arghavan and Grawitter, Josua and Stark, Holger},
  journal={Soft Matter},
  volume={20},
  number={8},
  pages={1800--1814},
  year={2024},
  publisher={Royal Society of Chemistry}
}

@article{partovifard2026controlled,
  title={Controlled route to active turbulence: Filling an activity spot with topological defects},
  author={Partovifard, Arghavan and Stark, Holger},
  journal={Soft Matter},
  year={2026},
  publisher={Royal Society of Chemistry}
}

@article{arneodo1998towards,
  title={Towards log-normal statistics in high Reynolds number turbulence},
  author={Arneodo, A and Manneville, S and Muzy, JF},
  journal={The European Physical Journal B-Condensed Matter and Complex Systems},
  volume={1},
  number={1},
  pages={129--140},
  year={1998},
  publisher={Springer}
}

@article{kolmogorov1962refinement,
  title={A refinement of previous hypotheses concerning the local structure of turbulence in a viscous incompressible fluid at high Reynolds number},
  author={Kolmogorov, Andrey Nikolaevich},
  journal={Journal of Fluid Mechanics},
  volume={13},
  number={1},
  pages={82--85},
  year={1962},
  publisher={Cambridge University Press}
}

@article{perlekar2026flocking,
  title={Flocking and mesoscale turbulence in three-dimensional active fluids},
  author={Perlekar, Prasad},
  journal={arXiv preprint arXiv:2601.17674},
  year={2026}
}

@article{rana2024defect,
  title={Defect turbulence in a dense suspension of polar, active swimmers},
  author={Rana, Navdeep and Chatterjee, Rayan and Ro, Sunghan and Levine, Dov and Ramaswamy, Sriram and Perlekar, Prasad},
  journal={Physical Review E},
  volume={109},
  number={2},
  pages={024603},
  year={2024},
  publisher={APS}
}

@article{jain2025instabilities,
  title={Instabilities and turbulence in extensile swimmer suspensions},
  author={Jain, Purnima and Rana, Navdeep and Benzi, Roberto and Perlekar, Prasad},
  journal={Physical Review Fluids},
  volume={10},
  number={11},
  pages={114602},
  year={2025},
  publisher={APS}
}

@article{padhan2025cahn,
  title={The Cahn--Hilliard--Navier--Stokes framework for multiphase fluid flows: laminar, turbulent and active},
  author={Padhan, Nadia Bihari and Pandit, Rahul},
  journal={Journal of Fluid Mechanics},
  volume={1010},
  pages={P1},
  year={2025},
  publisher={Cambridge University Press}
}

@article{jain2024inertia,
  title={Inertia drives concentration-wave turbulence in swimmer suspensions},
  author={Jain, Purnima and Rana, Navdeep and Ramaswamy, Sriram and Perlekar, Prasad},
  journal={Physical Review Letters},
  volume={133},
  number={15},
  pages={158302},
  year={2024},
  publisher={APS}
}

@article{bees2025emergent,
  title={Emergent asymmetry in confined bioconvection},
  author={Bees, Martin A and Perlekar, Prasad},
  journal={Journal of Fluid Mechanics},
  volume={1003},
  pages={A21},
  year={2025},
  publisher={Cambridge University Press}
}

@article{ramaswamy2019active,
  title={Active fluids},
  author={Ramaswamy, Sriram},
  journal={Nature Reviews Physics},
  volume={1},
  number={11},
  pages={640--642},
  year={2019},
  publisher={Nature Publishing Group UK London}
}

@article{aditi2002hydrodynamic,
  title={Hydrodynamic fluctuations and instabilities in ordered suspensions of self-propelled particles},
  author={Aditi Simha, R and Ramaswamy, Sriram},
  journal={Physical review letters},
  volume={89},
  number={5},
  pages={058101},
  year={2002},
  publisher={APS}
}

@article{doostmohammadi2018active,
  title={Active nematics},
  author={Doostmohammadi, Amin and Ign{\'e}s-Mullol, Jordi and Yeomans, Julia M and Sagu{\'e}s, Francesc},
  journal={Nature communications},
  volume={9},
  number={1},
  pages={3246},
  year={2018},
  publisher={Nature Publishing Group UK London}
}

@article{martinez2021scaling,
  title={Scaling regimes of active turbulence with external dissipation},
  author={Mart{\'\i}nez-Prat, Berta and Alert, Ricard and Meng, Fanlong and Ign{\'e}s-Mullol, Jordi and Joanny, Jean-Fran{\c{c}}ois and Casademunt, Jaume and Golestanian, Ramin and Sagu{\'e}s, Francesc},
  journal={Physical Review X},
  volume={11},
  number={3},
  pages={031065},
  year={2021},
  publisher={APS}
}

@article{peng2021imaging,
  title={Imaging the emergence of bacterial turbulence: Phase diagram and transition kinetics},
  author={Peng, Yi and Liu, Zhengyang and Cheng, Xiang},
  journal={Science advances},
  volume={7},
  number={17},
  pages={eabd1240},
  year={2021},
  publisher={American Association for the Advancement of Science}
}

@article{chakrabarti2024cytoplasmic,
  title={Cytoplasmic stirring by active carpets},
  author={Chakrabarti, Brato and Rachh, Manas and Shvartsman, Stanislav Y and Shelley, Michael J},
  journal={Proceedings of the National Academy of Sciences},
  volume={121},
  number={30},
  pages={e2405114121},
  year={2024},
  publisher={National Academy of Sciences}
}

@article{chakrabarti2023collective,
  title={Collective motion and pattern formation in phase-synchronizing active fluids},
  author={Chakrabarti, Brato and Shelley, Michael J and F{\"u}rthauer, Sebastian},
  journal={Physical Review Letters},
  volume={130},
  number={12},
  pages={128202},
  year={2023},
  publisher={APS}
}

@article{golestanian2011hydrodynamic,
  title={Hydrodynamic synchronization at low Reynolds number},
  author={Golestanian, Ramin and Yeomans, Julia M and Uchida, Nariya},
  journal={Soft Matter},
  volume={7},
  number={7},
  pages={3074--3082},
  year={2011},
  publisher={Royal Society of Chemistry}
}

@article{uchida2010synchronization,
  title={Synchronization and collective dynamics in a carpet of microfluidic rotors},
  author={Uchida, Nariya and Golestanian, Ramin},
  journal={Physical review letters},
  volume={104},
  number={17},
  pages={178103},
  year={2010},
  publisher={APS}
}

@article{mukherjee2024turbulent,
  title={Turbulent flows are not uniformly multifractal},
  author={Mukherjee, Siddhartha and Murugan, Sugan Durai and Mukherjee, Ritwik and Ray, Samriddhi Sankar},
  journal={Physical Review Letters},
  volume={132},
  number={18},
  pages={184002},
  year={2024},
  publisher={APS}
}

@BOOK{Frisch-Book,
  TITLE = {Turbulence: The Legacy of A. N. Kolmogorov},
  AUTHOR = {Frisch, U},
  YEAR = {1996},
  PUBLISHER = {Cambridge University Press},
}

@misc{Sahoo2025,
      title={Fluctuating interfaces in barotropic beta-plane turbulence}, 
      author={Sandip Sahoo and Samriddhi Sankar Ray},
      year={2025},
      eprint={2507.23493},
      archivePrefix={arXiv},
      primaryClass={physics.flu-dyn},
      url={https://arxiv.org/abs/2507.23493}, 
}

@article{Patteson2018,
	author = {Patteson, Alison E. and Gopinath, Arvind and Arratia, Paulo E.},
	da = {2018/12/18},
	doi = {10.1038/s41467-018-07781-y},
	id = {Patteson2018},
	isbn = {2041-1723},
	journal = {Nature Communications},
	number = {1},
	pages = {5373},
	title = {The propagation of active-passive interfaces in bacterial swarms},
	ty = {JOUR},
	url = {https://doi.org/10.1038/s41467-018-07781-y},
	volume = {9},
	year = {2018}
}

@article{Frisch-Parisi,
	author={"Frisch, U. and Parisi, G."},
	title={Turbulence and Predictability of Geophysical Fluid Dynamics and Climate Dynamics},
	journal={Proceedings of the International School of Physics Enrico Fermi, Course LXXXVIII, Varenna, 1985},
	publisher={North-Holland},
	year={1985},
	URL={https://cir.nii.ac.jp/crid/1570291225411264384}
}

@article{MS-PRL,
  title = {Simple multifractal cascade model for fully developed turbulence},
  author = {Meneveau, C. and Sreenivasan, K. R.},
  journal = {Phys. Rev. Lett.},
  volume = {59},
  issue = {13},
  pages = {1424--1427},
  numpages = {0},
  year = {1987},
  month = {Sep},
  publisher = {American Physical Society},
  doi = {10.1103/PhysRevLett.59.1424},
  url = {https://link.aps.org/doi/10.1103/PhysRevLett.59.1424}
}

@article{MS-JFM, title={The multifractal nature of turbulent energy dissipation}, volume={224}, DOI={10.1017/S0022112091001830}, journal={Journal of Fluid Mechanics}, publisher={Cambridge University Press}, author={Meneveau, Charles and Sreenivasan, K. R.}, year={1991}, pages={429–484}}

@article{MS-Nucl,
title = {The multifractal spectrum of the dissipation field in turbulent flows},
journal = {Nuclear Physics B - Proceedings Supplements},
volume = {2},
pages = {49-76},
year = {1987},
issn = {0920-5632},
doi = {https://doi.org/10.1016/0920-5632(87)90008-9},
url = {https://www.sciencedirect.com/science/article/pii/0920563287900089},
author = {C. Meneveau and K.R. Sreenivasan}
}

@article{saghatchi2022nematic,
  title={Nematic order condensation and topological defects in inertial active nematics},
  author={Saghatchi, Roozbeh and Yildiz, Mehmet and Doostmohammadi, Amin},
  journal={Physical Review E},
  volume={106},
  number={1},
  pages={014705},
  year={2022},
  publisher={APS}
}

@article{linkmann2019phase,
  title={Phase transition to large scale coherent structures in two-dimensional active matter turbulence},
  author={Linkmann, Moritz and Boffetta, Guido and Marchetti, M Cristina and Eckhardt, Bruno},
  journal={Physical review letters},
  volume={122},
  number={21},
  pages={214503},
  year={2019},
  publisher={APS}
}

@article{li2022topological,
  title={Topological transitions, turbulent-like motion and long-time-tails driven by cell division in biological tissues},
  author={Li, Xin and Sinha, Sumit and Kirkpatrick, Theodore R and Thirumalai, Dave},
  journal={arXiv preprint arXiv:2211.14410},
  year={2022}
}

@article{sokolov2012physical,
  title={Physical properties of collective motion in suspensions of bacteria},
  author={Sokolov, Andrey and Aranson, Igor S},
  journal={Physical review letters},
  volume={109},
  number={24},
  pages={248109},
  year={2012},
  publisher={APS}
}

@article{mazzino2021unraveling,
  title={Unraveling the secrets of turbulence in a fluid puff},
  author={Mazzino, Andrea and Rosti, Marco Edoardo},
  journal={Physical Review Letters},
  volume={127},
  number={9},
  pages={094501},
  year={2021},
  publisher={APS}
}

@article{matsuzawa2023creation,
  title={Creation of an isolated turbulent blob fed by vortex rings},
  author={Matsuzawa, Takumi and Mitchell, Noah P and Perrard, St{\'e}phane and Irvine, William TM},
  journal={Nature Physics},
  volume={19},
  number={8},
  pages={1193--1200},
  year={2023},
  publisher={Nature Publishing Group UK London}
}

@article{joy2020friction,
  title={Friction scaling laws for transport in active turbulence},
  author={CP, Sanjay and Joy, Ashwin},
  journal={Physical Review Fluids},
  volume={5},
  number={2},
  pages={024302},
  year={2020},
  publisher={APS}
}

@article{martinez2025interfacial,
  title={Interfacial morphodynamics of proliferating microbial communities},
  author={Mart{\'\i}nez-Calvo, Alejandro and Trenado-Yuste, Carolina and Lee, Hyunseok and Gore, Jeff and Wingreen, Ned S and Datta, Sujit S},
  journal={Physical Review X},
  volume={15},
  number={1},
  pages={011016},
  year={2025},
  publisher={APS}
}

@article{nair2021lagrangian,
  title={A Lagrangian study of interfaces at the edges of cumulus clouds},
  author={Nair, Vishnu and Heus, Thijs and van Reeuwijk, Maarten},
  journal={Journal of the Atmospheric Sciences},
  volume={78},
  number={8},
  pages={2397--2412},
  year={2021}
}

@article{hokmabad2025spatial,
  title={Spatial self-organization of confined bacterial suspensions},
  author={Hokmabad, Babak Vajdi and Mart{\'\i}nez-Calvo, Alejandro and Gonzalez La Corte, Sebastian and Datta, Sujit S},
  journal={Proceedings of the National Academy of Sciences},
  volume={122},
  number={41},
  pages={e2503983122},
  year={2025},
  publisher={National Academy of Sciences}
}

@article{lemma2023spatio,
  title={Spatio-temporal patterning of extensile active stresses in microtubule-based active fluids},
  author={Lemma, Linnea M and Varghese, Minu and Ross, Tyler D and Thomson, Matt and Baskaran, Aparna and Dogic, Zvonimir},
  journal={PNAS nexus},
  volume={2},
  number={5},
  pages={pgad130},
  year={2023},
  publisher={Oxford University Press US}
}

@article{heus2008subsiding,
  title={Subsiding shells around shallow cumulus clouds},
  author={Heus, Thijs and Jonker, Harm JJ},
  journal={Journal of the Atmospheric Sciences},
  volume={65},
  number={3},
  pages={1003--1018},
  year={2008}
}

@article{chamkor-chaudhuri-natcomm,
	author = {Singh, Chamkor and Chaudhuri, Abhishek},
	da = {2024/05/02},
	doi = {10.1038/s41467-024-47727-1},
	id = {Singh2024},
	isbn = {2041-1723},
	journal = {Nature Communications},
	number = {1},
	pages = {3704},
	title = {Anomalous dynamics of a passive droplet in active turbulence},
	ty = {JOUR},
	volume = {15},
	year = {2024},
}

@article{arlt2019dynamics,
  title={Dynamics-dependent density distribution in active suspensions},
  author={Arlt, Jochen and Martinez, Vincent A and Dawson, Angela and Pilizota, Teuta and Poon, Wilson CK},
  journal={Nature communications},
  volume={10},
  number={1},
  pages={2321},
  year={2019},
  publisher={Nature Publishing Group UK London}
}

@article{xin2000front,
  title={Front propagation in heterogeneous media},
  author={Xin, Jack},
  journal={SIAM review},
  volume={42},
  number={2},
  pages={161--230},
  year={2000},
  publisher={SIAM}
}

@article{pocheau1994scale,
  title={Scale invariance in turbulent front propagation},
  author={Pocheau, A},
  journal={Physical Review E},
  volume={49},
  number={2},
  pages={1109},
  year={1994},
  publisher={APS}
}

@article{roy2023small,
  title={Small-scale intermittency of premixed turbulent flames},
  author={Roy, Amitesh and Picardo, Jason R and Emerson, Benjamin and Lieuwen, Tim C and Sujith, Raman I},
  journal={Journal of Fluid Mechanics},
  volume={957},
  pages={A21},
  year={2023}
}

@article{koudella2004reaction,
  title={Reaction front propagation in a turbulent flow},
  author={Koudella, Christophe R and Neufeld, Zolt{\'a}n},
  journal={Physical Review E—Statistical, Nonlinear, and Soft Matter Physics},
  volume={70},
  number={2},
  pages={026307},
  year={2004},
  publisher={APS}
}

@article{corwin2012kardar,
  title={The Kardar--Parisi--Zhang equation and universality class},
  author={Corwin, Ivan},
  journal={Random matrices: Theory and applications},
  volume={1},
  number={01},
  pages={1130001},
  year={2012},
  publisher={World Scientific}
}

@article{bentkamp2022statistical,
  title={The statistical geometry of material loops in turbulence},
  author={Bentkamp, Lukas and Drivas, Theodore D and Lalescu, Cristian C and Wilczek, Michael},
  journal={Nature Communications},
  volume={13},
  number={1},
  pages={2088},
  year={2022},
  publisher={Nature Publishing Group UK London}
}

@book{shiffman2024nature,
  title={The Nature of Code: Simulating Natural Systems with JavaScript},
  author={Shiffman, Daniel},
  year={2024},
  publisher={No Starch Press}
}

@book{pearson2011generative,
  title={Generative art: a practical guide using processing},
  author={Pearson, Matt},
  year={2011},
  publisher={Simon and Schuster}
}

@book{reas2007processing,
  title={Processing: a programming handbook for visual designers and artists},
  author={Reas, Casey and Fry, Ben},
  year={2007},
  publisher={Mit Press}
}

@article{chauhan2014turbulent,
  title={The turbulent/non-turbulent interface and entrainment in a boundary layer},
  author={Chauhan, Kapil and Philip, Jimmy and De Silva, Charitha M and Hutchins, Nicholas and Marusic, Ivan},
  journal={Journal of Fluid Mechanics},
  volume={742},
  pages={119--151},
  year={2014},
  publisher={Cambridge University Press}
}

@article{borrell2016properties,
  title={Properties of the turbulent/non-turbulent interface in boundary layers},
  author={Borrell, Guillem and Jim{\'e}nez, Javier},
  journal={Journal of Fluid Mechanics},
  volume={801},
  pages={554--596},
  year={2016},
  publisher={Cambridge University Press}
}

@article{mukherjeelocalPRL2024,
  title = {Turbulent Flows Are Not Uniformly Multifractal},
  author = {Mukherjee, Siddhartha and Murugan, Sugan Durai and Mukherjee, Ritwik and Ray, Samriddhi Sankar},
  journal = {Phys. Rev. Lett.},
  volume = {132},
  issue = {18},
  pages = {184002},
  numpages = {6},
  year = {2024},
  month = {May},
  publisher = {American Physical Society},
  doi = {10.1103/PhysRevLett.132.184002},
}

@article{westerweel2009momentum,
  title={Momentum and scalar transport at the turbulent/non-turbulent interface of a jet},
  author={Westerweel, Jerry and Fukushima, C and Pedersen, Jakob Martin and Hunt, Julian CR},
  journal={Journal of Fluid Mechanics},
  volume={631},
  pages={199--230},
  year={2009},
  publisher={Cambridge University Press}
}

@article{elsinga2019turbulent,
  title={How the turbulent/non-turbulent interface is different from internal turbulence},
  author={Elsinga, GE and Da Silva, CB},
  journal={Journal of Fluid Mechanics},
  volume={866},
  pages={216--238},
  year={2019},
  publisher={Cambridge University Press}
}

@article{watanabe2015turbulent,
  title={Turbulent mixing of passive scalar near turbulent and non-turbulent interface in mixing layers},
  author={Watanabe, T and Sakai, Y and Nagata, K and Ito, Y and Hayase, T},
  journal={Physics of Fluids},
  volume={27},
  number={8},
  year={2015},
  publisher={AIP Publishing}
}

@article{kashyap2025emergence,
  title={Emergence of local ordering and giant number fluctuations in active turbulence},
  author={Kashyap, Kirti and Kiran, Kolluru Venkata and Gupta, Anupam},
  journal={arXiv preprint arXiv:2507.04890},
  year={2025}
}

@article{grassberger1983characterization,
  title={Characterization of strange attractors},
  author={Grassberger, Peter and Procaccia, Itamar},
  journal={Physical review letters},
  volume={50},
  number={5},
  pages={346},
  year={1983},
  doi={https://doi.org/10.1103/PhysRevLett.50.346},
  publisher={APS}
}

@article{puggioni2023flocking,
  title={Flocking turbulence of microswimmers in confined domains},
  author={Puggioni, Leonardo and Boffetta, Guido and Musacchio, Stefano},
  journal={Physical Review E},
  volume={107},
  number={5},
  pages={055107},
  year={2023},
  publisher={APS}
}

@article{pal2016binary,
  title={Binary-fluid turbulence: Signatures of multifractal droplet dynamics and dissipation reduction},
  author={Pal, Nairita and Perlekar, Prasad and Gupta, Anupam and Pandit, Rahul},
  journal={Physical Review E},
  volume={93},
  number={6},
  pages={063115},
  year={2016},
  publisher={APS}
}

@article{meneveau1987simple,
  title={Simple multifractal cascade model for fully developed turbulence},
  author={Meneveau, C and Sreenivasan, KR},
  journal={Physical review letters},
  volume={59},
  number={13},
  pages={1424},
  year={1987},
  publisher={APS}
}

@article{meneveau1991multifractal,
  title={The multifractal nature of turbulent energy dissipation},
  author={Meneveau, Charles and Sreenivasan, KR},
  journal={Journal of Fluid Mechanics},
  volume={224},
  pages={429--484},
  year={1991},
  publisher={Cambridge University Press}
}

@book{frisch1995turbulence,
  title={Turbulence: the legacy of AN Kolmogorov},
  author={Frisch, Uriel},
  year={1995},
  publisher={Cambridge university press}
}

@article{boffetta2012two,
  title={Two-dimensional turbulence},
  author={Boffetta, Guido and Ecke, Robert E},
  journal={Annual review of fluid mechanics},
  volume={44},
  number={1},
  pages={427--451},
  year={2012},
  publisher={Annual Reviews}
}

@article{cao1999statistics,
  title={Statistics and structures of pressure in isotropic turbulence},
  author={Cao, Nianzheng and Chen, Shiyi and Doolen, Gary D},
  journal={Physics of Fluids},
  volume={11},
  number={8},
  pages={2235--2250},
  year={1999},
  publisher={American Institute of Physics}
}

@article{pumir1994numerical,
  title={A numerical study of pressure fluctuations in three-dimensional, incompressible, homogeneous, isotropic turbulence},
  author={Pumir, Alain},
  journal={Physics of Fluids},
  volume={6},
  number={6},
  pages={2071--2083},
  year={1994},
  publisher={American Institute of Physics}
}

@article{padhan2023activity,
  title={Activity-induced droplet propulsion and multifractality},
  author={Padhan, Nadia Bihari and Pandit, Rahul},
  journal={Physical Review Research},
  volume={5},
  number={3},
  pages={L032013},
  year={2023},
  publisher={APS}
}

@article{persistencePRL2011,
  title = {Persistence Problem in Two-Dimensional Fluid Turbulence},
  author = {Perlekar, Prasad and Ray, Samriddhi Sankar and Mitra, Dhrubaditya and Pandit, Rahul},
  journal = {Phys. Rev. Lett.},
  volume = {106},
  issue = {5},
  pages = {054501},
  numpages = {4},
  year = {2011},
  month = {Feb},
  publisher = {American Physical Society},
  doi = {10.1103/PhysRevLett.106.054501},
}

@article{3Dpersistence,
  title = {How long do particles spend in vortical regions in turbulent flows?},
  author = {Bhatnagar, Akshay and Gupta, Anupam and Mitra, Dhrubaditya and Pandit, Rahul and Perlekar, Prasad},
  journal = {Phys. Rev. E},
  volume = {94},
  issue = {5},
  pages = {053119},
  numpages = {8},
  year = {2016},
  month = {Nov},
  publisher = {American Physical Society},
  doi = {10.1103/PhysRevE.94.053119},
}

@article{active-persistence,
  title = {Persistence in active turbulence},
  author = {Manoharan, Amal and CP, Sanjay and Joy, Ashwin},
  journal = {Phys. Rev. E},
  volume = {108},
  issue = {6},
  pages = {L062602},
  numpages = {5},
  year = {2023},
  month = {Dec},
  publisher = {American Physical Society},
  doi = {10.1103/PhysRevE.108.L062602},
}

@article{marchetti2013hydrodynamics,
  title={Hydrodynamics of soft active matter},
  author={Marchetti, M Cristina and Joanny, Jean-Fran{\c{c}}ois and Ramaswamy, Sriram and Liverpool, Tanniemola B and Prost, Jacques and Rao, Madan and Simha, R Aditi},
  journal={Reviews of modern physics},
  volume={85},
  number={3},
  pages={1143--1189},
  year={2013},
  publisher={APS}
}

@article{liu2021viscoelastic,
  title={Viscoelastic control of spatiotemporal order in bacterial active matter},
  author={Liu, Song and Shankar, Suraj and Marchetti, M Cristina and Wu, Yilin},
  journal={Nature},
  volume={590},
  number={7844},
  pages={80--84},
  year={2021},
  publisher={Nature Publishing Group UK London}
}

@article{schuppler2016boundaries,
  title={Boundaries steer the contraction of active gels},
  author={Schuppler, Matthias and Keber, Felix C and Kr{\"o}ger, Martin and Bausch, Andreas R},
  journal={Nature communications},
  volume={7},
  number={1},
  pages={13120},
  year={2016},
  publisher={Nature Publishing Group UK London}
}

@article{arlt2018painting,
  title={Painting with light-powered bacteria},
  author={Arlt, Jochen and Martinez, Vincent A and Dawson, Angela and Pilizota, Teuta and Poon, Wilson CK},
  journal={Nature communications},
  volume={9},
  number={1},
  pages={768},
  year={2018},
  publisher={Nature Publishing Group UK London}
}

@article{alexakis2023far,
  title={How far does turbulence spread?},
  author={Alexakis, Alexandros},
  journal={Journal of Fluid Mechanics},
  volume={977},
  pages={R1},
  year={2023},
  publisher={Cambridge University Press}
}

@article{puggioni2022giant,
  title={Giant vortex dynamics in confined bacterial turbulence},
  author={Puggioni, L and Boffetta, G and Musacchio, S},
  journal={Physical Review E},
  volume={106},
  number={5},
  pages={055103},
  year={2022},
  publisher={APS}
}

@article{wioland2016directed,
  title={Directed collective motion of bacteria under channel confinement},
  author={Wioland, Hugo and Lushi, Enkeleida and Goldstein, Raymond E},
  journal={New Journal of Physics},
  volume={18},
  number={7},
  pages={075002},
  year={2016},
  publisher={IOP Publishing}
}

@article{reinken2020organizing,
  title={Organizing bacterial vortex lattices by periodic obstacle arrays},
  author={Reinken, Henning and Nishiguchi, Daiki and Heidenreich, Sebastian and Sokolov, Andrey and B{\"a}r, Markus and Klapp, Sabine HL and Aranson, Igor S},
  journal={Communications Physics},
  volume={3},
  number={1},
  pages={76},
  year={2020},
  publisher={Nature Publishing Group UK London}
}

@article{nishiguchi2024vortex,
  title={Vortex reversal is a precursor of confined bacterial turbulence},
  author={Nishiguchi, Daiki and Shiratani, Sora and Takeuchi, Kazumasa A and Aranson, Igor S},
  journal={arXiv preprint arXiv:2407.05269},
  year={2024}
}

@article{dunkel2013fluid,
  title={Fluid dynamics of bacterial turbulence},
  author={Dunkel, J{\"o}rn and Heidenreich, Sebastian and Drescher, Knut and Wensink, Henricus H and B{\"a}r, Markus and Goldstein, Raymond E},
  journal={Physical review letters},
  volume={110},
  number={22},
  pages={228102},
  year={2013},
  publisher={APS}
}

@article{reinken2022ising,
  title={Ising-like critical behavior of vortex lattices in an active fluid},
  author={Reinken, Henning and Heidenreich, Sebastian and B{\"a}r, Markus and Klapp, Sabine HL},
  journal={Physical Review Letters},
  volume={128},
  number={4},
  pages={048004},
  year={2022},
  publisher={APS}
}

@article{nishiguchi2018engineering,
  title={Engineering bacterial vortex lattice via direct laser lithography},
  author={Nishiguchi, Daiki and Aranson, Igor S and Snezhko, Alexey and Sokolov, Andrey},
  journal={Nature communications},
  volume={9},
  number={1},
  pages={4486},
  year={2018},
  publisher={Nature Publishing Group UK London}
}

@article{james2018turbulence,
  title={Turbulence and turbulent pattern formation in a minimal model for active fluids},
  author={James, Martin and Bos, Wouter JT and Wilczek, Michael},
  journal={Physical Review Fluids},
  volume={3},
  number={6},
  pages={061101},
  year={2018},
  publisher={APS}
}

@article{kumar2014flocking,
  title={Flocking at a distance in active granular matter},
  author={Kumar, Nitin and Soni, Harsh and Ramaswamy, Sriram and Sood, AK},
  journal={Nature communications},
  volume={5},
  number={1},
  pages={4688},
  year={2014},
  publisher={Nature Publishing Group UK London}
}

@article{james2018vortex,
  title={Vortex dynamics and Lagrangian statistics in a model for active turbulence},
  author={James, Martin and Wilczek, Michael},
  journal={The European Physical Journal E},
  volume={41},
  pages={1--6},
  year={2018},
  publisher={Springer}
}

@article{martinez2019selection,
  title={Selection mechanism at the onset of active turbulence},
  author={Mart{\'\i}nez-Prat, Berta and Ign{\'e}s-Mullol, Jordi and Casademunt, Jaume and Sagu{\'e}s, Francesc},
  journal={Nature physics},
  volume={15},
  number={4},
  pages={362--366},
  year={2019},
  publisher={Nature Publishing Group UK London}
}

@article{bratanov2015new,
  title={New class of turbulence in active fluids},
  author={Bratanov, Vasil and Jenko, Frank and Frey, Erwin},
  journal={Proceedings of the National Academy of Sciences},
  volume={112},
  number={49},
  pages={15048--15053},
  year={2015},
  publisher={National Acad Sciences}
}

@article{aranson2022bacterial,
  title={Bacterial active matter},
  author={Aranson, Igor S},
  journal={Reports on Progress in Physics},
  volume={85},
  number={7},
  pages={076601},
  year={2022},
  publisher={IOP Publishing}
}

@article{petroff2015fast,
  title={Fast-moving bacteria self-organize into active two-dimensional crystals of rotating cells},
  author={Petroff, Alexander P and Wu, Xiao-Lun and Libchaber, Albert},
  journal={Physical review letters},
  volume={114},
  number={15},
  pages={158102},
  year={2015},
  publisher={APS}
}

@article{engelhardt2022novel,
  title={Novel form of collective movement by soil bacteria},
  author={Engelhardt, IC and Patko, Daniel and Liu, Y and Mimault, M and de Las Heras Martinez, G and George, TS and MacDonald, M and Ptashnyk, M and Sukhodub, T and Stanley-Wall, NR and others},
  journal={The ISME Journal},
  volume={16},
  number={10},
  pages={2337--2347},
  year={2022},
  publisher={Oxford University Press}
}

@article{be2019statistical,
  title={A statistical physics view of swarming bacteria},
  author={Be’er, Avraham and Ariel, Gil},
  journal={Movement ecology},
  volume={7},
  pages={1--17},
  year={2019},
  publisher={Springer}
}

@article{zhang2021spatiotemporal,
  title={Spatiotemporal control of liquid crystal structure and dynamics through activity patterning},
  author={Zhang, Rui and Redford, Steven A and Ruijgrok, Paul V and Kumar, Nitin and Mozaffari, Ali and Zemsky, Sasha and Dinner, Aaron R and Vitelli, Vincenzo and Bryant, Zev and Gardel, Margaret L and others},
  journal={Nature materials},
  volume={20},
  number={6},
  pages={875--882},
  year={2021},
  publisher={Nature Publishing Group UK London}
}

@article{palacci2013living,
  title={Living crystals of light-activated colloidal surfers},
  author={Palacci, Jeremie and Sacanna, Stefano and Steinberg, Asher Preska and Pine, David J and Chaikin, Paul M},
  journal={Science},
  volume={339},
  number={6122},
  pages={936--940},
  year={2013},
  publisher={American Association for the Advancement of Science}
}

@article{frangipane2018dynamic,
  title={Dynamic density shaping of photokinetic E. coli},
  author={Frangipane, Giacomo and Dell'Arciprete, Dario and Petracchini, Serena and Maggi, Claudio and Saglimbeni, Filippo and Bianchi, Silvio and Vizsnyiczai, Gaszton and Bernardini, Maria Lina and Di Leonardo, Roberto},
  journal={Elife},
  volume={7},
  pages={e36608},
  year={2018},
  publisher={eLife Sciences Publications, Ltd}
}

@article{ross2019controlling,
  title={Controlling organization and forces in active matter through optically defined boundaries},
  author={Ross, Tyler D and Lee, Heun Jin and Qu, Zijie and Banks, Rachel A and Phillips, Rob and Thomson, Matt},
  journal={Nature},
  volume={572},
  number={7768},
  pages={224--229},
  year={2019},
  publisher={Nature Publishing Group UK London}
}

@article{shankar2022topological,
  title={Topological active matter},
  author={Shankar, Suraj and Souslov, Anton and Bowick, Mark J and Marchetti, M Cristina and Vitelli, Vincenzo},
  journal={Nature Reviews Physics},
  volume={4},
  number={6},
  pages={380--398},
  year={2022},
  publisher={Nature Publishing Group UK London}
}

@article{zhang2021autonomous,
  title={Autonomous materials systems from active liquid crystals},
  author={Zhang, Rui and Mozaffari, Ali and de Pablo, Juan J},
  journal={Nature Reviews Materials},
  volume={6},
  number={5},
  pages={437--453},
  year={2021},
  publisher={Nature Publishing Group UK London}
}

@article{shankar2019hydrodynamics,
  title={Hydrodynamics of active defects: From order to chaos to defect ordering},
  author={Shankar, Suraj and Marchetti, M Cristina},
  journal={Physical Review X},
  volume={9},
  number={4},
  pages={041047},
  year={2019},
  publisher={APS}
}

@article{wensink2012meso,
  title={Meso-scale turbulence in living fluids},
  author={Wensink, Henricus H and Dunkel, J{\"o}rn and Heidenreich, Sebastian and Drescher, Knut and Goldstein, Raymond E and L{\"o}wen, Hartmut and Yeomans, Julia M},
  journal={Proceedings of the national academy of sciences},
  volume={109},
  number={36},
  pages={14308--14313},
  year={2012},
  publisher={National Acad Sciences}
}

@article{cp2020friction,
  title={Friction scaling laws for transport in active turbulence},
  author={CP, Sanjay and Joy, Ashwin},
  journal={Physical Review Fluids},
  volume={5},
  number={2},
  pages={024302},
  year={2020},
  publisher={APS}
}

@article{kiran2024onset,
  title = {Onset of Intermittency and Multiscaling in Active Turbulence},
  author = {Kiran, Kolluru Venkata and Kumar, Kunal and Gupta, Anupam and Pandit, Rahul and Ray, Samriddhi Sankar},
  journal = {Phys. Rev. Lett.},
  volume = {134},
  issue = {8},
  pages = {088302},
  numpages = {6},
  year = {2025},
  month = {Feb},
  publisher = {American Physical Society},
  doi = {10.1103/PhysRevLett.134.088302},
}

@article{alert2022active,
  title={Active turbulence},
  author={Alert, Ricard and Casademunt, Jaume and Joanny, Jean-Fran{\c{c}}ois},
  journal={Annual Review of Condensed Matter Physics},
  volume={13},
  number={1},
  pages={143--170},
  year={2022},
  publisher={Annual Reviews}
}

@article{singh2022lagrangian,
  title={Lagrangian manifestation of anomalies in active turbulence},
  author={Singh, Rahul K and Mukherjee, Siddhartha and Ray, Samriddhi Sankar},
  journal={Physical Review Fluids},
  volume={7},
  number={3},
  pages={033101},
  year={2022},
  publisher={APS}
}

@article{mukherjee2021anomalous,
  title={Anomalous diffusion and L{\'e}vy walks distinguish active from inertial turbulence},
  author={Mukherjee, Siddhartha and Singh, Rahul K and James, Martin and Ray, Samriddhi Sankar},
  journal={Physical Review Letters},
  volume={127},
  number={11},
  pages={118001},
  year={2021},
  publisher={APS}
}

@article{mukherjee2023intermittency,
  title={Intermittency, fluctuations and maximal chaos in an emergent universal state of active turbulence},
  author={Mukherjee, Siddhartha and Singh, Rahul K and James, Martin and Ray, Samriddhi Sankar},
  journal={Nature Physics},
  volume={19},
  number={6},
  pages={891--897},
  year={2023},
  publisher={Nature Publishing Group UK London}
}

@article{doostmohammadi2017onset,
  title={Onset of meso-scale turbulence in active nematics},
  author={Doostmohammadi, Amin and Shendruk, Tyler N and Thijssen, Kristian and Yeomans, Julia M},
  journal={Nature communications},
  volume={8},
  number={1},
  pages={15326},
  year={2017},
  publisher={Nature Publishing Group UK London}
}

@article{thampi2014active,
  title={Active nematic materials with substrate friction},
  author={Thampi, Sumesh P and Golestanian, Ramin and Yeomans, Julia M},
  journal={Physical Review E},
  volume={90},
  number={6},
  pages={062307},
  year={2014},
  publisher={APS}
}

@article{doostmohammadi2016stabilization,
  title={Stabilization of active matter by flow-vortex lattices and defect ordering},
  author={Doostmohammadi, Amin and Adamer, Michael F and Thampi, Sumesh P and Yeomans, Julia M},
  journal={Nature communications},
  volume={7},
  number={1},
  pages={10557},
  year={2016},
  publisher={Nature Publishing Group UK London}
}

@article{toner1998flocks,
  title={Flocks, herds, and schools: A quantitative theory of flocking},
  author={Toner, John and Tu, Yuhai},
  journal={Physical review E},
  volume={58},
  number={4},
  pages={4828},
  year={1998},
  publisher={APS}
}

@article{chandragiri2020flow,
  title={Flow states and transitions of an active nematic in a three-dimensional channel},
  author={Chandragiri, Santhan and Doostmohammadi, Amin and Yeomans, Julia M and Thampi, Sumesh P},
  journal={Physical Review Letters},
  volume={125},
  number={14},
  pages={148002},
  year={2020},
  publisher={APS}
}

@article{natan2022mixed,
  title={Mixed-species bacterial swarms show an interplay of mixing and segregation across scales},
  author={Natan, Gal and Worlitzer, Vasco M and Ariel, Gil and Be’er, Avraham},
  journal={Scientific Reports},
  volume={12},
  number={1},
  pages={16500},
  year={2022},
  publisher={Nature Publishing Group UK London}
}

@article{ariel2015swarming,
  title={Swarming bacteria migrate by L{\'e}vy Walk},
  author={Ariel, Gil and Rabani, Amit and Benisty, Sivan and Partridge, Jonathan D and Harshey, Rasika M and Be'Er, Avraham},
  journal={Nature communications},
  volume={6},
  number={1},
  pages={8396},
  year={2015},
  publisher={Nature Publishing Group UK London}
}

@article{rorai2022coexistence,
  title={Coexistence of active and hydrodynamic turbulence in two-dimensional active nematics},
  author={Rorai, Cecilia and Toschi, Federico and Pagonabarraga, Ignacio},
  journal={Physical Review Letters},
  volume={129},
  number={21},
  pages={218001},
  year={2022},
  publisher={APS}
}

@article{toner2005hydrodynamics,
  title={Hydrodynamics and phases of flocks},
  author={Toner, John and Tu, Yuhai and Ramaswamy, Sriram},
  journal={Annals of Physics},
  volume={318},
  number={1},
  pages={170--244},
  year={2005},
  publisher={Elsevier}
}

@article{nishiguchi2015mesoscopic,
  title={Mesoscopic turbulence and local order in Janus particles self-propelling under an ac electric field},
  author={Nishiguchi, Daiki and Sano, Masaki},
  journal={Physical Review E},
  volume={92},
  number={5},
  pages={052309},
  year={2015},
  publisher={APS}
}

@article{ginot2018aggregation,
  title={Aggregation-fragmentation and individual dynamics of active clusters},
  author={Ginot, F{\'e}lix and Theurkauff, Isaac and Detcheverry, F and Ybert, Christophe and Cottin-Bizonne, C{\'e}cile},
  journal={Nature communications},
  volume={9},
  number={1},
  pages={696},
  year={2018},
  publisher={Nature Publishing Group UK London}
}

@article{colin2021multiple,
  title={Multiple functions of flagellar motility and chemotaxis in bacterial physiology},
  author={Colin, Remy and Ni, Bin and Laganenka, Leanid and Sourjik, Victor},
  journal={FEMS microbiology reviews},
  volume={45},
  number={6},
  pages={fuab038},
  year={2021},
  publisher={Oxford University Press}
}

@article{chen2017weak,
  title={Weak synchronization and large-scale collective oscillation in dense bacterial suspensions},
  author={Chen, Chong and Liu, Song and Shi, Xia-qing and Chat{\'e}, Hugues and Wu, Yilin},
  journal={Nature},
  volume={542},
  number={7640},
  pages={210--214},
  year={2017},
  publisher={Nature Publishing Group UK London}
}

@article{butler2010cell,
  title={Cell density and mobility protect swarming bacteria against antibiotics},
  author={Butler, Mitchell T and Wang, Qingfeng and Harshey, Rasika M},
  journal={Proceedings of the National Academy of Sciences},
  volume={107},
  number={8},
  pages={3776--3781},
  year={2010},
  publisher={National Acad Sciences}
}

@article{lai2009swarming,
  title={Swarming motility: a multicellular behaviour conferring antimicrobial resistance},
  author={Lai, Sandra and Tremblay, Julien and D{\'e}ziel, Eric},
  journal={Environmental microbiology},
  volume={11},
  number={1},
  pages={126--136},
  year={2009},
  publisher={Wiley Online Library}
}

@article{yang2019quenching,
  title={Quenching active swarms: effects of light exposure on collective motility in swarming Serratia marcescens},
  author={Yang, Junyi and Arratia, Paulo E and Patteson, Alison E and Gopinath, Arvind},
  journal={Journal of the Royal Society Interface},
  volume={16},
  number={156},
  pages={20180960},
  year={2019},
  publisher={The Royal Society}
}

@article{james2021emergence,
  title={Emergence and melting of active vortex crystals},
  author={James, Martin and Suchla, Dominik Anton and Dunkel, J{\"o}rn and Wilczek, Michael},
  journal={Nature communications},
  volume={12},
  number={1},
  pages={5630},
  year={2021},
  publisher={Nature Publishing Group UK London}
}

@article{sanchez2012spontaneous,
  title={Spontaneous motion in hierarchically assembled active matter},
  author={Sanchez, Tim and Chen, Daniel TN and DeCamp, Stephen J and Heymann, Michael and Dogic, Zvonimir},
  journal={Nature},
  volume={491},
  number={7424},
  pages={431--434},
  year={2012},
  publisher={Nature Publishing Group UK London}
}

@article{gautam2024harnessing,
  title={Harnessing density to control the duration of intermittent L{\'e}vy walks in bacterial turbulence},
  author={Gautam, Dhananjay and Meena, Hemlata and Matheshwaran, Saravanan and Chandran, Sivasurender},
  journal={Physical Review E},
  volume={110},
  number={1},
  pages={L012601},
  year={2024},
  publisher={APS}
}

@article{hoff2009prokaryotic,
  title={Prokaryotic phototaxis},
  author={Hoff, Wouter D and van der Horst, Michael A and Nudel, Clara B and Hellingwerf, Klaas J},
  journal={Chemotaxis: Methods and Protocols},
  pages={25--49},
  year={2009},
  publisher={Springer}
}

@article{wu2017transition,
  title={Transition from turbulent to coherent flows in confined three-dimensional active fluids},
  author={Wu, Kun-Ta and Hishamunda, Jean Bernard and Chen, Daniel TN and DeCamp, Stephen J and Chang, Ya-Wen and Fern{\'a}ndez-Nieves, Alberto and Fraden, Seth and Dogic, Zvonimir},
  journal={Science},
  volume={355},
  number={6331},
  pages={eaal1979},
  year={2017},
  publisher={American Association for the Advancement of Science}
}

@article{thar2003bacteria,
  title={Bacteria are not too small for spatial sensing of chemical gradients: an experimental evidence},
  author={Thar, Roland and K{\"u}hl, Michael},
  journal={Proceedings of the National Academy of Sciences},
  volume={100},
  number={10},
  pages={5748--5753},
  year={2003},
  publisher={National Acad Sciences}
}

@article{wilde2017light,
  title={Light-controlled motility in prokaryotes and the problem of directional light perception},
  author={Wilde, Annegret and Mullineaux, Conrad W},
  journal={FEMS microbiology reviews},
  volume={41},
  number={6},
  pages={900--922},
  year={2017},
  publisher={Oxford University Press}
}

@Misc{omega-YT-strip,
  author       = {},
  howpublished = {See \url{https://www.youtube.com/shorts/AzAhfmAAwyc} for an animation showing the evolution of the vorticity field 
  for a striped activity pattern.},
}

@Misc{omega-giant,
  author       = {},
  howpublished = {See \url{https://youtube.com/shorts/AIcsM2GP3d4} for an animation showing the formation of a large vortex, similar to what is obtained for highly active turbulence under circular confinement.},
}

@Misc{omega-YT-circle,
  author       = {},
  howpublished = {See animation showing the development of the vorticity field 
  for an activity quench over a circular geometry \url{https://www.youtube.com/watch?v=VJJg_SMyfw8}, along with a composite comparison of the emerging interface \url{https://www.youtube.com/shorts/a1vjBuiQSLM}.},
}

@Misc{interface-YT-circle,
  author       = {},
  howpublished = {See \url{https://www.youtube.com/shorts/l6AVRfRtIyo} for an animation showing the evolution of the interface separating the 
	  highly active flow from a region low activity.},
}

\section*{\textbf{Acknowledgements}}
	SM would like to thank Jason Picardo for discussions on this problem.	
	SSR acknowledges  the Indo–French
	Centre for the Promotion of Advanced Scientific Research
	(IFCPAR/CEFIPRA, project no. 6704-1) for support. SS and SSR acknowledge support of the Department of Atomic Energy, Government of India, under project no. RTI4019 The simulations were
	performed on the ICTS clusters Mario, Tetris, and Contra. RM and SSR
	acknowledge the support of the DAE, Government of India, under projects
	nos.  12-R\&D-TFR-5.10-1100 and RTI4019.	
	SM acknowledges the IITK Initiation Grant project IITK/ME/2024316 and the Govt. of India grant ANRF/ECRG/2024/002467/ENS.  This research was supported in part by the
	International Centre for Theoretical Sciences (ICTS) for participating
	in the programs ---  \textit{Field Theory and Turbulence}
	(code: ICTS/ftt2023/12) and \textit{Turbulence: Problems at the
	Interface of Mathematics and Physics} (code: ICTS/TPIMP2020/12). SSR thanks 
	the Isaac Newton Institute for Mathematical Sciences,
	Cambridge, for support and hospitality during the programme
	\textit{Anti-diffusive dynamics: from sub-cellular to astrophysical
	scales} (EPSRC grant EP/R014604/1), where part of the work on this
	paper was undertaken. 
\section*{\textbf{Author contributions}}
SM and SSR conceived and supervised the research. The Eulerian calculations and analysis were performed by SM and the 
Lagrangian calculations and analysis were performed by KK. The manuscript was written together by all the authors.

\section*{\textbf{Competing interests}}
The authors declare no competing interests.

  \end{document}